\documentclass[a4paper,onecolumn,11pt,accepted=2023-07-18]{quantumarticle}%twocolumn
\pdfoutput=1

\usepackage[utf8]{inputenc}
\usepackage[english]{babel}
\usepackage[T1]{fontenc}
\usepackage{amsmath}
\usepackage{hyperref}
\usepackage{amsthm}

\usepackage[numbers,sort&compress]{natbib}

\usepackage{graphicx}% Include figure files
\usepackage{dcolumn}% Align table columns on decimal point
\usepackage{amssymb}
\usepackage{mathtools}
\usepackage{float}
\usepackage{bm}

\usepackage{color,amsfonts}%amsmath,
\usepackage[usenames,dvipsnames]{xcolor}

\usepackage[normalem]{ulem}
\usepackage{textcomp}

\newcommand{\xfrac}[2]{{#1}/{#2}}
\newcommand{\rfrac}[2]{\left({#1}/{#2}\right)}

\newcommand{\beq}{\begin{equation}}
\newcommand{\eeq}{\end{equation}}
\newcommand{\bqa}{\begin{eqnarray}}
\newcommand{\eqa}{\end{eqnarray}}

\newcommand{\erf}[1]{Eq.~(\ref{#1})}

\newcommand{\ket}[1]{\left|{#1}\right\rangle}

\newcommand{\sch}{Schr\"odinger}

\newcommand{\ie}{{\em i.e.}}
\newcommand{\eg}{{\em e.g.}}
\newcommand{\etc}{{\em etc.}}
\newcommand{\etal}{{\em et al.}}

\newcommand{\vt}{$^\vartheta$}

\newtheorem{theorem}{Theorem}

\begin{document}

\title[A ``thoughtful'' Local Friendliness no-go theorem]{A ``thoughtful'' Local Friendliness no-go theorem:  
 a prospective experiment with new assumptions to suit}
\author{Howard M. Wiseman}
\affiliation{Centre for Quantum Computation and Communication Technology (Australian Research Council), \\ Centre for Quantum Dynamics, Griffith University, Yuggera Country, 
Brisbane, Queensland 4111, Australia}
\author{Eric G. Cavalcanti}
\affiliation{Centre for Quantum Dynamics, Griffith University, Yugambeh Country, 
Gold Coast, Queensland 4222, Australia}
\author{Eleanor G. Rieffel}
\affiliation{QuAIL (Quantum Artifical Intelligence Laboratory), NASA Ames Research Center, Moffett Field, CA 94035, United States of America}

\begin{abstract}
A recent paper by two of us and co-workers~\cite{CQD20}, based on an extended Wigner's friend scenario, demonstrated that certain empirical correlations predicted by quantum theory (QT) violate inequalities derived from a set of metaphysical assumptions we called ``Local Friendliness'' (LF). These assumptions are strictly weaker than those used for deriving Bell inequalities. Crucial to the theorem was the premise that a quantum system with reversible evolution could be an observer (colloquially, a ``friend''). However, that paper was noncommittal on what would constitute an observer  for the purpose of an experiment.  Here, we present a new LF no-go theorem which takes seriously the idea that a system's having  {\em thoughts} is a sufficient condition for it to be an observer. Our new derivation of the LF inequalities uses four metaphysical assumptions, three of which are thought-related, including one that is explicitly called  ``Friendliness''. These four assumptions, in conjunction, %imply 
allow one to derive LF inequalities for experiments involving the type of system that ``Friendliness'' refers to.  %commits one to accepting as an observer. 
In addition to these four metaphysical assumptions, this new no-go theorem  requires two assumptions about what is {\em technologically} feasible: Human-Level Artificial Intelligence, and Universal Quantum Computing which is fast and large scale. The latter is often motivated by the belief that QT is universal, but this is {\em not} an assumption of the theorem. The intent of the new theorem is to give a clear goal for future experimentalists, and a clear motivation for trying to achieve that goal, by using assumptions that are (i) logically independent, (ii) widely held, (iii) not reliant on the exact correctness of QT, and (iv) relevant to how different interpretations or modifications of QT respond to the no-go theorem. The popular stance that ``quantum theory needs no interpretation'' does not question any of our assumptions and so is ruled out by the theorem. Finally, we quantitatively discuss how difficult the experiment we envisage would be, and briefly discuss milestones on the paths towards it.
\end{abstract}

\maketitle

\newpage
\tableofcontents

\newpage

\section{Introduction} \label{sec:intro}

Wigner introduced  his now famous ``friend'' scenario  in 1961~\cite{Wigner61} to present what he saw as an incompatibility between quantum theory (QT) and what one might call, in a loose sense, {\em friendliness}: % (as per the italics added below): 
\begin{quotation}
It is natural to ask about the situation if one does not make [an] observation oneself but lets someone else carry it out. What is the wave function if my friend looked [at the quantum object]? The answer is that  
	\dots~one could [only] attribute a wave function to the joint system [of] friend plus object, $\alpha (\psi_1 \times \chi_1) + \beta (\psi_2 \times \chi_2)$ [which] follows from the linearity of the equations.  [This] appears absurd because it implies that my friend was in a state of suspended animation before he answered my question [about what he saw]. 
	
	It follows that the being with a consciousness must have a different role in quantum mechanics than %[{\em sic.}] 
	the inanimate measuring device. In particular, the quantum mechanical equations of motion cannot be linear if [it] is accepted \dots\ that ``my friend'' has the same types of impressions and sensations as I. \dots\ To deny the existence of the consciousness of a friend %to this extent 
	\dots\ is surely an unnatural attitude, approaching solipsism, and few people, in their hearts, will go along with it.
\end{quotation}

Wigner's argument was not a theorem with the same sort of rigour as, for example, Bell's soon-to-appear theorem~\cite{Bell64}. Recently, however, there has been a surge of renewed interest in the Wigner's friend scenario, due to the recognition that it is possible to obtain theorems by combining it with the setup (separated entangled systems and measurement choices) and formalism of Bell's theorem~\cite{BruknerBook,BruknerLF, FR18, Proietti19,Baumann18,Healey2018,Bau19,CQD20}. In particular, inspired by Brukner~\cite{BruknerBook,BruknerLF}, two of us and co-workers~\cite{CQD20} introduced the concept {\em Local Friendliness} (LF). This was defined in Ref.~\cite{CQD20} as the conjunction of three metaphysical assumptions\footnote{By ``metaphysical assumptions'' we mean fundamental assumptions that may be satisfied or violated by a physical theory, or by an ``interpretation'' of a physical theory, broadly construed.}: 
{\em No Super\-determinism} (permitting free choices), {\em Locality} (also known as ``parameter independence''), and {\em Absoluteness of Observed Events} (as opposed to such events being relative to an observer or ``world''). From this conjunction, certain inequalities, which we called LF inequalities, were derived, which would be violated under the technological assumption that arbitrary quantum operations could be performed on quantum systems considered to be observers. Thus, if this technological assumption holds, then LF must be false, a result which may be called a LF no-go theorem\footnote{The theorem was obtained by three of the authors: An{\'\i}bal Utreras-Alarc{\'o}n, Eric Cavalcanti, and Howard Wiseman. The fact that LF inequalities are, in general, strictly weaker (\ie, harder to violate) than Bell inequalities was proven together with the fourth theory author, Yeong-Cherng Liang.}~\cite{CQD20}. 

As well as proving the LF no-go theorem, Bong {\em et al.}~\cite{CQD20} presented a ``proof-of-principle'' experiment, in which various LF inequalities were violated. In this experiment, the role of a friend --- an observer to whom one of the ``observed events'' in the theorem pertains --- was played by a binary degree-of-freedom of a single photon (its path). The paper readily admitted that such a friend ``would not typically be considered a macroscopic, sentient observer as originally envisioned by Wigner.'' However, the paper was intentionally  noncommittal about what could be considered an adequate ``friend'' or ``observer''. That is, the ``observer'' in the theorem of Ref.~\cite{CQD20} should be interpreted as a placeholder to be defined by a candidate theory or ideology. For example, if every physical system can be considered as an observer, as claimed by Refs.~\cite{Proietti19} and~\cite{Rovelli20}, then the Wigner+Bell experiments performed with single-qubit friends~\cite{Proietti19,CQD20} already rule out LF\footnote{The first of these experiments, Ref.~\cite{Proietti19}, was published prior to the LF no-go theorem of Ref.~\cite{CQD20}, and so it is only a demonstration of the violation of LF inequalities in hindsight.}. But the majority of physicists and philosophers would surely not be convinced that these experiments disprove LF.

The fact that current experiments have only a proof-of-principle status is the essential motivation for the current paper. But addressing it requires not only  thinking about a new type of experiment. It also requires formulating a new (though closely related) theorem to suit, and to motivate, that type of experiment. We lay out, and then unfold, these two goals in the next subsection.

\subsection{The goals of this paper}
\label{sec:goals}

The first goal of this paper is to come up with a feasible, though ambitious, proposal for an experiment that would be broadly convincing as a test of LF. The second goal is to modify the assumptions of the LF no-go theorem of Ref.~\cite{CQD20} so as to obtain a new theorem of experimental metaphysics. The new assumptions are carefully chosen to attempt to motivate our experimental colleagues (or, more realistically, future generations of experimentalists) to perform the associated experiment. 
Specifically, the two goals are:
\begin{enumerate}
    \item To make the proposed experiment satisfy the following three desiderata: 
\begin{enumerate}
    \item That the experiment be {\em plausibly} achievable, if present technological trends continue, in a matter of decades rather than centuries.
    \item That, according to standard QT, the experiment should yield correlations violating a LF inequality. 
    \item That such a violation (if occurring) convince the largest possible part of the interested community that LF is false. 
    \label{item:LargePart}
\end{enumerate}

    \item To use assumptions in the associated theorem that are natural in the following senses: 
    \begin{enumerate} 
    \item They are logically independent of one another and of the assumption of the universal validity of QT; \label{item:logicindepQT} 
    \item They are relevant to many approaches to QT; \label{item:approaches} 
    \item They are widely held amongst physicists and other scientists; 
    \item \label{item:noFreedomForSkeptics}
They include a general principle of {\em friendliness}, in the same spirit to which Wigner originally appealed,
so as to leave no freedom to the skeptic to avoid the conclusion of the theorem by excluding certain types of ``friends'' by fiat. 
    \end{enumerate}
\end{enumerate} 

In goal (\ref{item:LargePart}) we say ``the largest possible part'' of the interested community, rather than ``all'', because we expect that there would be some physicists or philosophers who would believe that LF (as defined in Ref.~\cite{CQD20}) {\em must} be true. Such scholars would presumably say that any experiment purporting to violate it must involve a ``friend'' which does not deserve observer status. This will be discussed more in Sec.~\ref{sec:IntroTheorem}, and a particular approach to QT (Spontaneous Collapse) that harmonizes with this attitude is reviewed in Sec.~\ref{sec:SCTs}. At this stage, we remind the reader that we have already stated our goal (\ref{item:noFreedomForSkeptics}) of how to deal with skeptics with this sort of attitude--- they  would have to reject a principle of {\em friendliness} that is general, precise, and explicitly stated (which was not the case in Ref.~\cite{CQD20}). 

We should also clarify that in goal (\ref{item:approaches}) we use the word ``approaches'' to QT to  include two types of theories. {\em Interpretations} in the strict sense may add (\eg\ hidden variables) to, or subtract (\eg\ a special role for measurement) from, the standard quantum formalism, but do not (or at least claim not to) alter its usual empirical predictions. {\em Modifications} are, in principle, empirically distinguishable from QT, in some regime not yet probed by experiment. Representative examples of both types will be given in Sec.~\ref{sec:app}.

The enumeration of the goals above is not meant to imply that the theorem we will present is the only fruitful way of formulating new theorems based upon that of Ref.~\cite{CQD20}. Moreover, the experiment we propose would, we envisage, be the culmination of a long research program involving increasingly sophisticated ``friends''. But, now, in the early stages of scholarly discourse about LF no-go theorems, it seems useful to present the strongest theorem (\ie\ using the metaphysically weakest assumptions that are still intuitive and relevant) that we can currently devise, compatible with the goals stated above.

Our new formulation is, in fact, significantly closer 
to what Wigner originally discussed than is that in Ref.~\cite{CQD20}.  Specifically, our theorem (soon to be stated, in Sec.~\ref{sec:IntroTheorem})  considers a friend who appears to be able to communicate thoughts comparable to our own. We intentionally use the word ``thoughts'', rather than, as Wigner used, ``consciousness'', because the former seems easier to identify and less controversial --- we wish to avoid debates about what constitutes consciousness or even whether it exists. The connotations of the word ``consciousness'' may also make humans unwilling to ascribe it to a non-human intelligence. Similarly, we wish to avoid having to speculate as to whether a given intelligence has the ``same types of impressions and sensations''~\cite{Wigner61} as our own. 

\subsection{The new theorem}\label{sec:IntroTheorem}

To distinguish from the LF no-go theorem of Ref.~\cite{CQD20}, we will call the no-go theorem of this paper the LF\vt\ no-go theorem, where the $\vartheta$ superscript indicates it is a variation in which \emph{thoughts} play a central role. This can be seen in the formal statement of our new theorem:
\begin{theorem}[LF\vt\ no-go] 
\end{theorem} \vspace{-2ex} \begin{flushleft}There is a contradiction between the conjunction of four metaphysical assumptions:\end{flushleft} 
\begin{enumerate}
\item {\sc Local Agency} \label{item:LA}
\item {\sc Physical Supervenience} \label{item:PS}
\item {\sc Ego Absolutism} \label{item:EA}
\item {\sc Friendliness} \label{item:F}
\end{enumerate}
(which together constitute LF\vt); and these two technological assumptions:
\begin{enumerate} \setcounter{enumi}{4}
\item {\sc Human-Level Artificial Intelligence} (HLAI) 
\label{item:HLAI}
\item {\sc Universal Quantum Computing} (UQC). \label{item:UQC}
\end{enumerate}
We detail each of these assumptions in the body of the paper.

Note that the universal validity of QT is {\em not} part of the theorem here, unlike the theorems of Refs.~\cite{BruknerLF,FR18}. Of course the second technological assumption is often motivated by a belief in the universal validity of QT (UVQT). The reader, recalling our goal (\ref{item:logicindepQT}) in Sec.~\ref{sec:intro} to use assumptions that are logically independent of the assumption of UVQT, may thus wonder whether we have missed the mark in assumption \ref{item:UQC}. However, when we say that the UQC assumption is independent of UVQT, we mean 
that it is logically possible, at present, to reject UQC and accept UVQT or
to reject UVQT and accept UQC (or reject both or accept both). See Sec.~\ref{sec:UQC} for further discussion on this point. 

It is hopefully obvious that QT, whether universally valid or not, plays {\em no} role in the metaphysical assumptions (\ref{item:LA}.)--(\ref{item:F}.). The Wigner's friend paradox (in a broad sense) is often presented, as Wigner did in the opening quote, as a clash between Wigner's {\em description} of his friend using QT, and the friend's own beliefs. But our theorem does not require Alice (playing the role of Wigner) to assign a quantum state to her friend, and in fact does not presume that Alice would have sufficient knowledge to assign a pure quantum state to her friend even if she wanted to and it was valid to do so. 

The experimentally testable inequalities which can be derived from LF\vt\ (the conjunction of the four metaphysical assumptions of our new theorem) are of identical form to the inequalities derived in Ref.~\cite{CQD20}. This is because, for a suitable experiment, LF\vt\ gives the same constraints on correlations as does the assumptions of LF, but where the application of LF requires the acceptance of the ``friend'' system as an observer. The interest in the new theorem lies in the assumptions behind it and the nature of the associated experiment. The new assumptions are complete; they remove the need (as discussed in the preceding subsection) for an additional implicit assumption that such-and-such a system in a particular experiment is an observer. Thus, because the proposed experiment is designed to fit the new assumptions, if the experiment violates a LF inequality, then it must be accepted that LF\vt\ %(the conjunction of the above metaphysical assumptions) 
is false. This is so even though, as discussed above, a skeptic may maintain that LF in the original sense of Ref.~\cite{CQD20} remains inviolate, by denying that the friend in the experiment we design is a system capable of making an observation.

\subsection{The structure of the paper}

We begin by recapitulating the LF no-go theorem of Ref.~\cite{CQD20} in Sec.~\ref{sec:TOLFNGT}. We present the proof explicitly in a way that makes it easy to compare with the LF\vt\ no-go theorem, and expand upon some issues of relevance for both theorems. 
Even readers who are familiar with Ref.~\cite{CQD20} are advised to read this section, especially Sec.~\ref{sec:Qviolations}, for its discussion of the disjoint roles of the metaphysical (theory-independent) assumptions and the technological (and quantum-theory-related) assumption in the LF no-go theorem. This discussion also applies, with some alterations, to our new theorem.  
The four metaphysical assumptions of the new theorem, whose conjunction is LF\vt, are defined and discussed in Sec.~\ref{sec:assM}. The two technological assumptions, along with the nature of the experiment that would be required to demonstrate the falsity of LF\vt, are presented in Sec.~\ref{sec:assE}. That section also discusses minor technological assumptions, plus ethical and other methodological requirements for the experiment. In Sec.~\ref{sec:app} we demonstrate the relevance of all six assumptions by briefly presenting six different approaches (or classes of approaches) to QT, each one violating one --- and arguably only one --- of the assumptions.  In Sec.~\ref{sec:exp} we estimate the quantum computing resources required to carry out the proposed experiment. We conclude in Sec.~\ref{sec:disc} with a discussion of future work, and of the most important current implication for our theorem: the stance that ``quantum theory needs no interpretation'', popularized by Fuchs and Peres~\cite{FucPer2000}, does not deny any of our assumptions and so is ruled out.

\section{The original LF no-go theorem} \label{sec:TOLFNGT}

The LF no-go theorem of Ref.~\cite{CQD20} can be considered a theorem of {\em experimental metaphysics}. The phrase  here in italics was introduced by Shimony~\cite{Shi84} to refer to the field initiated by Bell's theorem.  The theorem is, like Bell's, a strict theorem of experimental metaphysics in the following sense. It is a proof that, given certain scientific and technological assumptions, it would be possible to do a well defined experiment that would plausibly give a result that rules out the conjunction of certain well defined metaphysical assumptions.

The theorem builds upon a thought experiment introduced by Brukner~\cite{BruknerBook,BruknerLF}, involving a combination of a standard Bell-type scenario with a Wigner's friend scenario, called an \emph{Extended Wigner's Friend Scenario} (EWFS). Brukner's EWFS considered two observers in spatially separated labs, Charlie and Debbie, who initially share a pair of entangled quantum systems. They are the ``friends'' of  two ``super\-observervers'', Alice and Bob respectively, who are capable of performing arbitrary quantum operations on the entire contents of their respective friends' lab. As shown in Ref.~\cite{CQD20}, a more parsimonious scenario, involving only one friend (Charlie), already allows for violation of LF inequalities. We will focus on this simpler scenario here.

\subsection{Derivation of LF inequalities in an extended Wigner's friend scenario}\label{sec:DLFoiEWFS}

Before proceeding, it is important to make a distinction between the LF no-go theorem, on one hand, and the derivation of LF inequalities on the other. The LF inequalities are derived in a \emph{theory-independent} manner, without making any assumption specifically about QT. The LF no-go theorem is a proof that, under certain technological assumptions, QT allows for the violation of LF inequalities.  To make the distinction clear, we will first summarise the derivation of LF inequalities, without any allusion to quantum systems or operations. 
We note that all of this is completely analogous to the distinction between Bell's theorem and the derivation of Bell inequalities. In particular, we use the term ``theory-independent'' in exactly the same way as it is used in the literature on Bell inequalities. That is, the metaphysical assumptions only apply for a theoretical framework which take as {\em a priori} notions such as space-time, physical events, interventions, and observations. But we stress that such a framework does not assume the absolute reality of observations, or other physical events, that may appear in the theorem. 

\begin{figure}
    \centering
    \includegraphics{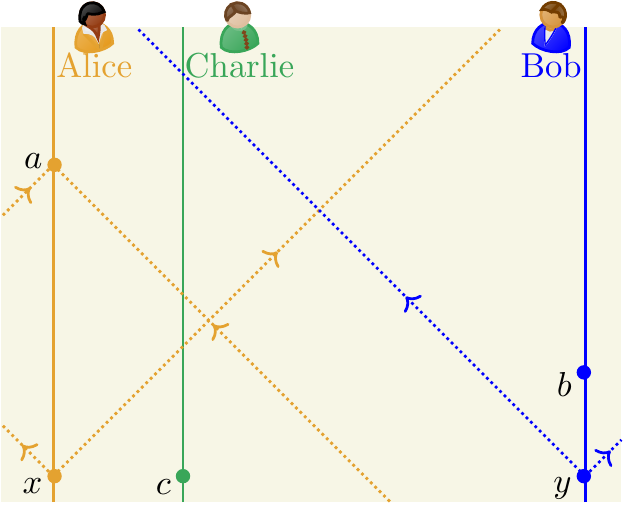}
    \caption{Space-time diagram of the events involved in the original LF no-go theorem. The solid vertical lines are the world-lines of the three parties (time increases upwards). Future-directed light-like lines (dotted) emanate from Alice's ($x$) and Bob's ($y$) freely chosen settings for their measurements, yielding outcomes $a$ and $b$ respectively. Past-directed light-like lines (dotted) converge on Alice's outcome $a$, showing that Charlie's outcome, $c$ (the result of a fixed measurement) is in the past light cone of $a$.  The application of {\sc Local Agency} relies on the fact that, of the outcomes $a$, $b$ and $c$,  only $a$ is in the future light-cone of $x$, and only $b$ is in the future light-cone of $y$. In the case $x=1$, Alice simply asks Charlie his outcome, and copies it so that $a=c$. In the case $x\neq 1$ she follows a far more complicated experimental procedure, involving Charlie and his lab, to obtain her outcome. }
    \label{fig:originalSTdiagram}
\end{figure}

The minimal scenario involves three observers, Charlie, Alice and Bob, as illustrated in Fig.~\ref{fig:originalSTdiagram}. In each run of the experiment, a bipartite system is prepared and distributed to Charlie and Bob in separate laboratories. Charlie performs a fixed measurement 
upon his subsystem, observing an outcome labelled by $c\in\{\pm 1\}$. Alice has a choice between two measurements labelled by $x\in\{1,2\}$, with outcomes labelled by $a\in\{\pm 1\}$. Bob has a choice between two measurements labelled by $y\in\{1,2\}$, with outcomes $b\in\{\pm 1\}$. Alice's measurement $x$ is chosen in an appropriately random fashion, within a space-time region not in the past light-cone of $c$ or $b$. Bob's measurement $y$ is also chosen in an appropriately random fashion, within a space-time region not in the past light-cone of $c$ or $a$. The experimental protocol says that, if Alice's measurement is $x=1$, she will ask Charlie for his observed outcome, and assign her outcome to be equal to that of Charlie\footnote{There is, of course, an implicit assumption that Charlie accurately reports his observed outcome. We will discuss the status of this assumption in Section \ref{sec:LFvsBell}.}, i.e.~$a=c$. 
Otherwise, she will perform a different measurement upon the contents of Charlie's lab (including Charlie). Repeating this experiment produces empirical probabilities $\wp(a,b|x,y)$ for the observations performed by Alice and Bob. Charlie's outcome $c$ is not included in these empirical probabilities, as it may be erased in the course of Alice's measurement when $x\neq 1$, as we will see when discussing the quantum realisation.

This is a complete description of the minimal scenario needed to derive an LF inequality. Note that we have not mentioned entangled quantum systems, nor did we specify that Alice is a super\-observerver to Charlie. Of course, those things will be needed to show that the LF inequalities can be violated by QT, but they are not needed for the derivation of the inequalities. With this scenario in mind, let us consider the constraints imposed by the LF assumption on $\wp(a,b|x,y)$.  The first metaphysical assumption that went into the LF assumption in Ref.~\cite{CQD20} is the following:

{\bf Absoluteness of Observed Events} (AOE): {\it An observed event is an absolute single event, and not relative to anything or anyone.}

This assumption implies that in each run of the experiment described above, the variables $a,b,c,x,y$ --- all of which are observed by some observer in every run --- take well defined, absolute, single values. The relative frequencies of these events define a probability distribution $P(a,b,c|x,y)$\footnote{Strictly, this requires that the system preparations in different runs of the experiment be interchangeable, which is  standard experimental assumption. A version of the LF no-go theorem without probabilistic assumptions was recently derived in~\cite{HadCav22}.}. The empirical probabilities $\wp(a,b|x,y)$ are related to the above by marginalisation. That is,  $\wp(a,b|x,y)=\sum_c P(a,b,c|x,y)$, for all $a,b,x,y$. We also have, from the scenario as constructed above, that $P(a|c,x=1,y)=\delta_{a,c}$, for all $a,c,y$.

Note that AOE is \emph{not} challenging Peres' dictum that ``unperformed experiments have no results''~\cite{Peres1978}. It is rather only an assumption that \emph{observed} events (and in particular, observed measurement outcomes) have absolute values. 

The LF no-go theorem of Ref.~\cite{CQD20} used two more metaphysical assumptions, {\em Locality} and {\em No-Super\-determinism}. This was a legacy from Brukner's theorem~\cite{BruknerLF}, which inspired it; as noted in  Ref.~\cite{CQD20}, the assumption of \emph{Local Agency} (introduced in Ref.~\cite{WisCav17}), implies both of those other assumptions. Here, we word it as follows:

{\bf Local Agency} (LA). {\it Any intervention, made in a manner appropriate for randomized experimental trials of a given phenomenon, is uncorrelated with any set of physical events that are relevant to that phenomenon and outside the future light-cone of that intervention.}

In our scenario, the \emph{interventions} assign values to the variables $x$ and $y$ for Alice's and Bob's choices of measurement. We borrow the term ``intervention'' from the literature on causal models~\cite{Pearl2000}, instead of the standard term ``free choice'' typically used in the context of Bell's theorem, to avoid the connotation that that these choices require a direct application of human ``free will''. In a given experimental situation, an intervention should be designed to make it as plausible as possible (within reason) that those choices are made via ``external variables'', \ie~variables that do not have causes among, or share common causes with, the other experimentally relevant variables. There is thus also a background assumption that for the purposes of any well defined experiment, it is possible to make interventions in a manner appropriate for randomized trials. But that is not an additional metaphysical assumption; it is a statement about the norms of scientific practice.  

As argued in Ref.~\cite{WisCav17}, LA  is more naturally formulated than both  Locality and No-Super\-determinism. For this reason, we use LA in the formulation of our new LF\vt\ no-go theorem (see Sec.~\ref{sec:IntroTheorem}). For ease of comparison, we will thus review the proof of the original LF no-go theorem using this assumption. For simplicity, we will still refer to the conjunction of AOE and LA as ``Local Friendliness'', which was the name used in Ref.~\cite{CQD20} for the conjunction of AOE, {Locality}, and {No-Super\-determinism}. 
 
Notwithstanding the above, it is certainly possible to split {Local Agency} into a natural assumption that concerns agency, and an  independent natural assumption that concerns locality {\em sensu lato}. For example, in the most recent systematization by two of us~\cite{CavWis21}, the two assumptions would be, respectively, {\em Interventionist Causation} and {\em Relativistic Causal Arrow} (and each of these can, in turn, be split into two further independent assumptions~\cite{CavWis21}). This brings out the essential role for relativity, and a forward-in-time causal arrow in  Local Agency. Thus, while we acknowledge Colbeck and Renner for coming up with the essential concept of LA in 2011~\cite{ColRen11}, our reasons for not adopting their term for it, {\em free choice}, are hopefully clear.

In the formulation of both AOE and LA, there is a background assumption that we can apply concepts such as ``light-cones'', ``future'', \etc, at least at the scale at, and in the theoretical context in which, humans perceive and interact with the world. It is also implicit in the definition of LA that a physical event takes place in some space-time volume bounded in all dimensions in some reference frame.

Given the existence of $P(a,b,c|x,y)$ implied by AOE, and the space-time relations between the variables in the scenario, as shown in Fig.~\ref{fig:originalSTdiagram}, LA implies (as does No-Super\-determinism) that $c$ is independent of $x$ and $y$, \ie,  $P(c|x,y)=P(c)$, for all $c,x,y$. It also implies (as does the assumption of {Locality}), that $P(a|c,x,y)=P(a|c,x)$ and $P(b|c,x,y)=P(b|c,y)$, for all values of the variables involved.

Following the method of proof of~\cite{CQD20}, it can be shown that these constraints imply the existence of a model with probabilities $P(abc|xy)$ satisfying 

\begin{equation}
\label{LFgeneral}
\wp(ab|xy)=\begin{cases}
\sum_{c}\delta_{a,c}P(b|cy)P(c) & \text{ if } x=1 \\ 
\sum_{c}P(ab|cxy)P(c) & \text{ if } x\neq1 \,,
\end{cases}
\end{equation}
where the only constraint on $P(ab|cxy)$ is that it satisfies Local Agency. This is equivalent to the constraint that, for each possible value of $c$, the distribution $P(ab|cxy)$ does not allow faster-than-light signalling between Alice and Bob. 

The results in~\cite{CQD20} imply that any Alice-Bob correlations $\wp(ab|xy)$ that have a model of the form of Eq.~\eqref{LFgeneral} must satisfy inequalities of the same form as the CHSH inequalities for this scenario, where Alice and Bob each have two settings with two outcomes\footnote{This can be seen as follows. In~\cite{CQD20}, all of the tight LF inequalities for the scenario involving 3 dichotomic measurements for each of Alice and Bob, and with a ``friend'' for each party (Charlie and Debbie) are derived. Among these facet inequalities are the ``semi-Brukner'' inequalities---CHSH-type inequalities involving only one friend. If we ignore Debbie (i.e.~ignore Bob's measurement $y=1$ in that scenario), and ignore Alice's measurement $x=3$, we recover the scenario considered here.}. In other words, all Alice-Bob Bell inequalities are mathematically equivalent to the LF inequalities, in this particular scenario (though {\em not} in scenarios including more than two settings for Alice~\cite{CQD20}). Therefore, violation of these inequalities in an Extended Wigner's Friend Scenario of the form described above implies the failure of Local Friendliness. 

Again we stress that the conclusion above has been reached without any reference to quantum theory. This is similar to how the observation of the violation of a Bell inequality in a standard Bell scenario implies the failure of any conjunction of assumptions that are sufficient to derive a Bell inequality, in a theory-independent manner. At this point, and still before any discussion of QT, we can see that the violation of Local Friendliness inequalities (in an EWFS) would have stronger implications than the violation of Bell inequalities\footnote{Some confusion may arise here from the fact that in the EWFS we consider, the LF inequalities have the same mathematical form as the Bell-CHSH inequalities. When we talk of a ``violation of a Bell inequality'', we mean a violation of such an inequality within a standard Bell-type scenario. When the same inequality is implied, within an EWFS, by Local Friendliness alone, we refer to it as an ``LF inequality''.}. This is because the derivation of LF inequalities uses strictly weaker metaphysical assumptions than the derivation of Bell inequalities, as we now discuss.

\subsection{Local Friendliness assumptions are weaker than Bell's assumptions}
\label{sec:LFvsBell}

To understand that the LF assumptions are strictly weaker, three points are key. 

First, Local Agency does not imply the notion of \emph{Local Causality} which Bell used in the 1976 version of his theorem~\cite{Bel76}. Local Causality says that, if space-like events are correlated, there must exist common causes, in their common past light-cone, that render those correlated events conditionally uncorrelated --- the ``hidden variables'' in Bell's theorem. By contrast, no events or variables are postulated to exist in Local Agency; it merely says that correlations cannot exist between a certain type of event (an intervention $x$ in an experiment) and certain other types of events (physical events relevant to that experiment and not in $x$'s future light-cone).

Second, while {Local Agency} {\em does} imply the notion of \emph{Locality} which Bell used in the original (1964) version of his theorem~\cite{Bell64}, Bell also required another assumption there, \emph{Predetermination}. The LF no-go theorem does not make use of the assumption of {Predetermination}, or anything akin to it. As an aside, we remind the reader that Local Agency takes care of an assumption of No-Super\-determinism or ``free choice'', which Bell made implicitly in 1964 but later explicitly recognized the need for in all versions of his theorem. (See Refs.~\cite{Wis14b,WisRie15,WisRieCav16,WisCav17} on Bell's different assumptions in different papers.)

Third, a metaphysical assumption of Absoluteness of Observed Events, or something like it, is required in Bell's theorem to address what Kent in 2005~\cite{Kent2005} called the ``collapse locality loophole''. This name for the loophole is unfortunate, as it does not necessarily have anything to do with collapse\footnote{Kent chose the name because he was considering the loophole in the context of objective collapse theories. Such theories would specify, more or less, when and where a measurement result occurred, as a result of wavefunction collapse. This way of closing the loophole would be attractive in, as Kent imagines, a potential scientific future where it is agreed that objective collapse occurs, whether because of compelling experimental evidence or for compelling theoretical reasons. However, in the present situation where there is no consensus that collapse theories are correct, and indeed no consensus about the ontology of measurement results, the loophole can only be addressed (not closed!) by a metaphysical assumption, such as Absoluteness of Observed Events.  Different approaches to QT will, of course, assign different truth values to this assumption, and the truth value may also differ depending on how the assumption, and the approach, are understood. One purpose of our new theorem (see next section) is to remove any ambiguity in the assumption, and to try to force proponents of approaches to be clearer, by introducing a more fine-grained set of assumptions to address this loophole.}.
As Kent put it later~\cite{Kent2020}, ``This loophole arises because, while Bell experiments are supposed to demonstrate nonlocal correlations between measurement outcomes on spacelike separated systems, we do not know for sure where in space-time the relevant measurement outcomes actually arise.'' That is, Bell's theorem requires that measurement outcomes {\em actually arise} (at approximately the space-time locations we would expect); in other words, that observed events (in space-time) are absolute. In particular, the derivation of Bell inequalities is blocked if one allows the reality of different parties' records to depend on those records' becoming jointly knowable to a single party~\cite{Kent2020}. 

The assumption of AOE in Bell's theorem is very often ignored (although not in Ref.~\cite{WisCav17}, where the assumption of {\em Macroreality} serves the same purpose). In particular, all Bell tests to date implicitly assume that the measurement outcomes have been ``observed'', in an absolute sense, by the time a macroscopic record is created, for some meaning of that phrase. This is, {\em a priori}, a reasonable assumption, but in the context of a theorem where \emph{all} assumptions seem {\em a priori} reasonable, it is a substantial assumption. If quantum mechanics is universally valid, after all, there may be no objective threshold where a record becomes ``macroscopic'' or ``irreversible''. 

Furthermore, assuming a record {\em is} absolutely real from the moment ($t$, say) the observation is made until the moment ($t'$, say) it is reported to another party, that record must also be reliable: the report at time $t'$ must match the observation at time $t$. That is to say, for the Bell inequalities to be applicable to the experiment, we must trust that the computer records of measurement results correspond to the actual measurement results they are supposed to be a record of. This is also required in a LF experiment: we must trust that Charlie's observed outcome is accurately reported to Alice. Indeed, it is arguably a more substantive issue in LF experiments, and we note that some authors have considered relaxations of this assumption~\cite{XSN21,Moreno2022}, leading to more stringent inequalities. See also Sec.~\ref{subsec:MethodDesutch} for further discussion of this issue and how to address it.

\subsection{Quantum violations of LF inequalities}\label{sec:Qviolations}

To show that QT in principle allows for the violation of LF inequalities, we consider the following situation as set out in Ref.~\cite{CQD20}. Charlie and Bob are in possession of two qubits, $Q_A$ and $Q_B$, with associated Hilbert spaces $\mathcal{H}_{Q_A}$ and $\mathcal{H}_{Q_B}$, initially prepared in an entangled state. Bob performs one of two measurements on his qubit, depending on the value of $y$, observing outcome $b$. Charlie, who is in a well isolated ``lab''\footnote{In any realisation of this experiment where LF inequalities are violated, Charlie will not be a human, and  Charlie's ``lab'' will not be a room-sized laboratory, of course. Here by Charlie's lab we mean Charlie (whatever he is) plus whatever additional systems (\eg, a real or virtual environment) must also be controlled in order to reverse Charlie's measurement (see below).  In Sec.~\ref{sec:exp}, we consider the minimalist scenario, where Charlie himself is isolated, for a short time), and no additional environment is needed.}, performs a measurement on his qubit in a fixed basis, associated with an operator $Z_A$, observing outcome $c$.

It is now that the status of Alice as a so-called super\-observer enters crucially into the discussion, before even her result $a$ does. In order to obtain violations of LF inequalities, we do not take Charlie's measurement to induce a collapse of the quantum state (depending on the result $c$), even though measurement-induced collapse is part of ``standard QT'' as one would find in a text-book. Rather, we take the quantum state of Charlie's lab to evolve like that of any other isolated physical quantum system in standard QT. This is because we are considering, as worded in Theorem 1 of Bong \etal, what can be concluded 
\begin{quote}
    if a superobserver can perform arbitrary quantum
operations on an observer and its environment,
\end{quote}
where here Alice has the role of the super\-observer. A more apt name for this role in our protocol might  be a super-manipulator, 
but in either case, Alice can control the evolution of Charlie as a physical system like any other. That this requires ``super-powers'' of Alice is because typically we imagine Charlie is a rather large and complex system. This will become much more explicit with our new theorem, in Sec.~\ref{sec:assE}. 

Thus, in the below, we will assume that unitary quantum theory can be used to determine the evolution of the quantum state of Charlie's lab (including Charlie). This quantum state can then be used to calculate the expected distribution $\wp(ab|xy)$ of Alice's and Bob's outcomes, which can then be compared to the constraints (inequalities) that follow from LF (the conjunction of the metaphysical assumptions) as in Sec.~\ref{sec:DLFoiEWFS}. The variable $c$, Charlie's outcome, that went into the derivation of the inequalities is {\em not} to be looked for, or found, in the quantum state of Charlie's lab. Its absolute existence is a consequence of the metaphysical assumption AOE, which is completely independent of the predictions of QT here, and indeed (as the theorem says) in direct contradiction with those predictions when combined with the metaphysical assumption of LA. 

 We denote everything in Charlie's lab (including Charlie), except for $Q_A$, as system $F_A$, with Hilbert space $\mathcal{H}_{F_A}$, and assume that this is a well isolated system. According to standard QT, the quantum state of Charlie's lab during his observation of the qubit evolves according to a unitary $U_{Z_A}$ acting on $\mathcal{H}_{F_A}\otimes\mathcal{H}_{Q_A}$. When $x=1$, Alice asks Charlie what he saw. In this instance, Alice is treating Charlie as an observer on equal footing with herself (as in Wigner's original scenario), and accepts his answer as a true representation of his observation $c$. She assigns this as her own outcome, so we can say $a=c$. In operational terms, Alice has effectively measured  the qubit $Q_A$ in the same basis as Charlie. 
 
 When $x=2$, Alice acts quite  differently. She now treats Charlie's lab as a physical system, and forces its quantum state to evolve according to the inverse unitary $U^\dagger_{Z_A}$. This reverses the interaction between Charlie and the qubit, restoring the entanglement between qubit $Q_A$ and Bob's qubit $Q_B$. Alice then proceeds to measure the qubit directly, in a basis different from that used by Charlie. Note again that we are not required to say anything about the existence, or reality, of $c$, in our discussion of this evolution. We are, in this section, only describing how the experiment is performed and how to calculate  $\wp(ab|xy)$. 
 
 As shown in Ref.~\cite{CQD20}, there exist states and measurements such that this $\wp(ab|xy)$ will violate a CHSH-type inequality. Since Local Friendliness requires that these inequalities be satisfied, this demonstrates that QT in principle allows for the violation of Local Friendliness. This is the LF no-go theorem of Ref.~\cite{CQD20}. Of course, this theorem is explicitly (see earlier quote) making the (physical, not metaphysical) assumption that it is possible to make Charlie's lab well isolated, and to perform the ``undoing'' unitary $U^\dagger_{Z_A}$. How plausible {\em those} assumptions are depends on what type of system $F_A$ is.
 
 In Ref.~\cite{CQD20}, an experiment was reported in which $Q_A$ and $F_A$ were indeed well isolated and could have their interaction reversed. Specifically, the qubit $Q_A$ being observed was a photon polarisation qubit, while this photon’s path was ``deemed'' the  ``observer'', $F_A$. Many readers would, we suspect, be disinclined to accept this as a genuine observer, or the interaction (using a polarising beam-splitter) as yielding an observed event. That is why Ref.~\cite{CQD20} said the experiment therein was ``best
described as a proof-of-principle version of the EWFS.'' But the LF no-go theorem of Ref.~\cite{CQD20} does not provide a criterion for what kind of system should be deemed an observer, or what manner of event would constitute an observation. This leaves it up to debate whether any particular experiment is actually a test of LF. To resolve this issue --- without explicitly using the concepts of ``observer'' or ``observation'' or ``observed event'' --- is the primary motivation for our new theorem, to which we now return.

\section{Metaphysical Assumptions for the Thoughtful LF no-go Theorem}
\label{sec:assM}

In Sec.~\ref{sec:assM4} we present the four metaphysical assumptions that are required for our LF\vt\ no-go theorem. As stated in Sec.~\ref{sec:goals}, we believe that each of our four metaphysical assumptions is widely held. Moreover, we will present each of them so as to make it appear plausible. However, it is critical to note that we are {\em not} arguing that they are all true. That is because we think that the two technological assumptions to be presented in Sec.~\ref{sec:assE} are also plausibly true, individually and collectively. But the conjunction of all six assumptions is a contradiction; that is the theorem. As to {\em which} assumption is not true, we adopt a strictly neutral position. Given that one must be false, for each one it is reasonable to believe that it is false. Indeed, for each one there is an approach to QT that rejects it; see Sec.~\ref{sec:app}.

Three of the four metaphysical assumptions (the last three) relate to the question with which we finished the preceding section, namely how to address the lack of a criterion for something to be an observed event in the Absoluteness of Observed Events assumption used in Ref.~\cite{CQD20}. That is, the conjunction of assumptions (\ref{item:PS}.)--(\ref{item:F}.) together give a specific criterion for a certain type of event, which plays the same role of the friend's observation in the original LF scenario,  to be absolute. As already discussed in Sec.~\ref{sec:goals}, we do not claim that the answer given here is the only reasonable answer. However, we do suggest that our answer would be more widely acceptable than any other answer that would allow an experiment that could, plausibly, (a) be performed in the foreseeable future, and (b) violate a LF inequality. The sense in which this conjunction of three assumptions in the LF\vt\ no-go theorem is functionally equivalent to the assumption of Absoluteness of Observed Events in the original LF no-go theorem is set out in Sec.~\ref{sec:LFfCMA}.

\subsection{The four metaphysical assumptions}
\label{sec:assM4}

\begin{enumerate}
 \item {\sc Local Agency}. 
Any intervention, made in a manner appropriate for randomized experimental trials of a given phenomenon, is uncorrelated with any set of physical events that are relevant to that phenomenon and outside the future light-cone of that intervention. 
 \label{ass:LA}
\end{enumerate}

This, the first metaphysical assumption we use in the LF\vt\ no-go theorem, was already presented and discussed in Sec.~\ref{sec:TOLFNGT}. It is worth reiterating, however, that {\sc Local Agency} is not the same as Bell's 1976 concept of Local Causality.  Moreover, neither implies the other. Also, to obtain Bell inequalities in the situation of a regular Bell experiment (with space-like-separated parties, and treating their observations as {\em absolute}) neither of these concepts is sufficient. With {\sc Local Causality} one needs an assumption relating to interventions such as No-Super\-determinism~\cite{WisCav17}, while for {\sc Local Agency} one needs an additional assumption relating to outcomes such as Predetermination~\cite{WisCav17}. Needless to say, we do not make an assumption like Predetermination for the theorem here.

\begin{enumerate} \setcounter{enumi}{1}
\item {\sc Physical Supervenience}. Any thought supervenes upon some physical event(s) in the brain (or other information-processing unit as appropriate) which can thus be located within a bounded region in space-time.
 \end{enumerate}
 This is certainly a widely held belief amongst scientists. It is compatible with the monist assumption that a thought is {\em nothing but} a physical process, but does not require that. It is also important to note that the assumption of {\sc Physical Supervenience} does not presume or imply that a thought, or the associated physical event or process (chain of events), has any {\it absolute} reality (a concept explained under {\sc Ego Absolutism} below). That is, {\sc Physical Supervenience}  is independent of the remaining metaphysical assumptions. 
 
 The confinement of the physical events upon which thoughts supervene in space-time is necessary to be able to apply the first assumption, {\sc Local Agency}, to prove the theorem. The reader may wonder whether this is in conflict with certain types of mind-externalism~\cite{sep-content-externalism} which hold that the physical processes on which a thought supervenes are not restricted to the brain, or even body, of a particular party, but may include that party's environment and even other parties with which they are interacting. In fact we think there is no conflict, because of the types of thoughts that are relevant for the LF\vt\ no-go theorem (and indeed for Bell's theorem in Kent's version with human observations~\cite{Kent2020}; see Sec.~\ref{sec:LFvsBell}). That is because the thoughts we will consider are tied to observations of inputs, and are assumed to take place on a time scale of a second or less. Moreover, for the case of the ``friend'' in the LF\vt\ experiment we will consider, this party is prevented from interacting with the physical environment external to its information-processing unit for the time during which the relevant thought would exist, were it to exist.

\begin{enumerate} \setcounter{enumi}{2}
\item {\sc Ego Absolutism}. My communicable thoughts are {\em absolutely} real.
 \end{enumerate}
Here you, the reader, when assessing whether you believe this assumption, should, of course, read it to yourself verbatim, not interpreting it as being a statement about the reality of the thoughts of me (the first author, as it happens)\footnote{This paragraph dates from the early versions of this paper, when the current first author was the sole author. We decided to keep it in first-person singular because it  thereby matches the phrasing of this assumption, and because it expresses some personal views of the first author.}. The formulation of the above assumption in the  first-person singular is not frivolous. Like Wigner in 1961\footnote{\guillemotleft There are several reasons for the return, on the part of most physical scientists, to the spirit of Descartes's ``\emph{Cogito ergo sum},'' which recognizes the thought, that is, the mind, as primary.\guillemotright~\cite{Wigner61}}, I agree with Descartes~\cite{Des1641} in the sense that I am more sure of the reality of my own thoughts than of anything else. Quite possibly the same is true for you, reader, again {\em mutatis mutandis}. This does not, of course, not mean that {\sc Ego Absolutism} must be true. But if Descartes was on the right track, then {\sc Ego Absolutism} is a philosophically minimalist criterion for an absolute ontology.

The meaning of ``absolute'' in all of the above needs explanation (as flagged by its italicization in the definition of {\sc Ego Absolutism}). For my thoughts to be real in an absolute sense means that I do not have to qualify any statements about my thoughts as being relative to anyone or anything. For example, when I have a thought ``it's  alive!''~it is not the case that this is my thought only in {\em this world}, or that this thought's existence is a fact only {\em for me}, or a fact only {\em relative} to certain other thoughts or states or facts or systems. Rather, my thought  exists unconditionally. Or at least, by {\sc Ego Absolutism}, this is the case for thoughts that are communicable rather than, for instance, ephemeral impressions or feelings whose existence I may even doubt myself. Finally, we note that a thought can be absolutely real even if it corresponds to an incorrect statement. For example, my thought ``it is raining'' may be absolutely real even if it is not in fact raining.

\begin{enumerate} \setcounter{enumi}{3}
\item {\sc Friendliness}. \label{ass:Friendliness} If %an {\em independent party} 
a system displays {\em independent} cognitive ability at least on par with my own, then they are a {\em party} with {\em cognition} at least on par with my own, and any {\em thought} they communicate is {\em as real as} any communicable thought of my own.
\end{enumerate}
As in the preceding definition, italics are used to emphasize terms that would profit from further commentary. Beginning at the end, ``as real as'' does not mean ``absolutely real, just as''. Rather, it means ``having the same degree of reality as''. Thus, for you, the reader (again reading first person pronouns as referring to yourself), it could imply absolutely real, or relatively real, or not real at all, depending on what you believe about your own communicable thoughts. Next, the italicization of (one instance) of ``thoughts'' is so we (the authors) can take the opportunity to re\"emphasize that ``thoughts'' is not to be equated with ``conscious thoughts'' or ``qualia'' or ``sensations'' or ``experiences''\footnote{Note that we are not suggesting that thoughts have greater claim to be absolutely real than feelings, or qualia.  In particular we do not intend any imputation about what types of beings are worthy of moral consideration, or the plausibility of claims on that matter.}. We reject the equation because it may make the assumption of {\sc Friendliness} less plausible, or less well defined, for some scientists or philosophers. 

We maintain the same stance with our use of the word ``cognition''; we mean a process comprising thoughts, not necessarily (as some dictionary definitions may offer) also experiences or sensations. As the root of the word implies, cognition entails thought-processes representing, acquiring, and producing knowledge. For cognition on par with my own, this includes self-knowledge, such as the ability to remember one's own short-term thought-processes, and the ability to think about a particular topic when required (not necessarily to the exclusion of other thoughts). Thus, if a party like this undertakes to think communicable thoughts about something then, under the {\sc Friendliness} assumption, he\footnote{We use the masculine gender for the party under consideration for linguistic convenience, because he appears most frequently in conjunction with a super\-observerver of feminine gender.  This follows the pattern for Charlie and Alice already used, though eventually we will give Charlie a new name.} does have thoughts, they are thoughts of human-level sophistication, and they are as real as my own. (That is, unless he is lying; see Sec.~\ref{subsec:MethodDesutch} for how to deal with this possibility.)

Moving on to ``party'', we are following our usage in Sec.~\ref{sec:TOLFNGT}. That is, we are now explicitly adopting the attitude that a system with human-level cognition is a participant in the experiment rather than a mere object of study. This has methodological implications discussed in Sec.~\ref{subsec:MethodDesutch}. Finally, regarding ``independent'', for this party's cognitive ability to be independent 
it is sufficient that histhoughts are not known to us unless he chooses to reveal them. That is, while this party's thoughts may (of course) be influenced by how other parties interact with him, his thoughts are not directly ``implanted''.

Turning now to the plausibility of this assumption, each of the authors of this paper attributes the same degree of reality (or ontological status, one might say) to the thoughts of other people, as to his or her own, and we would guess that you, reader, have the same inclination. This is so whether we are interacting with someone face-to-face, via video-chat, or via text-chat, for example. In the last case we are often communicating with someone we have never met face-to-face, whose physical appearance, age, and gender, may be unknown. Yet we can, after sufficient communication, discern thoughts which we treat as equal to, and equally real as, our own. It is not hard to believe that, in the not too distant future, we  will be doing this when communicating with parties who are  artificial intelligences. In fact, most people are remarkably easily duped into accepting the cognitive equality of simulated intelligence~\cite{marcus2014}. You, most likely a scientifically or philosophically trained reader of this paper, will probably require considerably better evidence before accepting an artificial intelligence as displaying cognitive ability on par with your own.  But, hypothesizing that such evidence will one day exist --- see assumption 5 below --- the question you have to ask yourself is whether there is any justification for dropping the {\sc Friendly} attitude you take (we presume) towards fellow humans. 

It is important to note that while 
the criterion of displaying ``cognitive ability at least on par with my own'' would require open-ended communication with the party in question, it   
does not require my being present at the experiment to do this. This is essential because it is unlikely that I, or any readers of the paper in its year of publication, will be alive to see the experiment we envisage performed. I can trust that others will be able to make that judgement, when the time comes. 
The reader may wonder why we insist on a criterion akin to a Turing-test, rather than also, or instead, requiring something of the functioning of the party's processor. One reason is that there is no currently agreed upon functional criterion for cognition, 
and there may still not be when our experiment is performable. A second, even more important, reason is that  
we want to apply this criterion to a party whose processor is a quantum computer, and in particular in a situation where the quantum mechanical state is a superposition of very different logical states; see Sec.~\ref{sec:UQC}. This is a situation for which conventional definitions of functionality may be impossible to apply, because quantum information is  not able to be accessed without altering the future behaviour of the processor. 

Finally, we want to stress that assumption 4 does not imply that being able to display cognitive ability at least on par with one's own is a {\em necessary} condition for a party's thoughts to be granted the same ontological status as one's own thoughts. As its wording implies, the assumption is only a sufficient condition.

\subsection{LF\vt\ from the conjunction of the metaphysical assumptions} \label{sec:LFfCMA}

We call the collection of all four of the above metaphysical assumptions LF\vt. For a suitable spatial arrangement of people, or other parties of human-level intelligence, and temporal arrangement of communications and controls, LF\vt\ implies the same constraints on correlations as in Ref.~\cite{CQD20}, \ie, a so-called LF polytope, as we now show. 

\begin{figure}
    \centering 
    \includegraphics{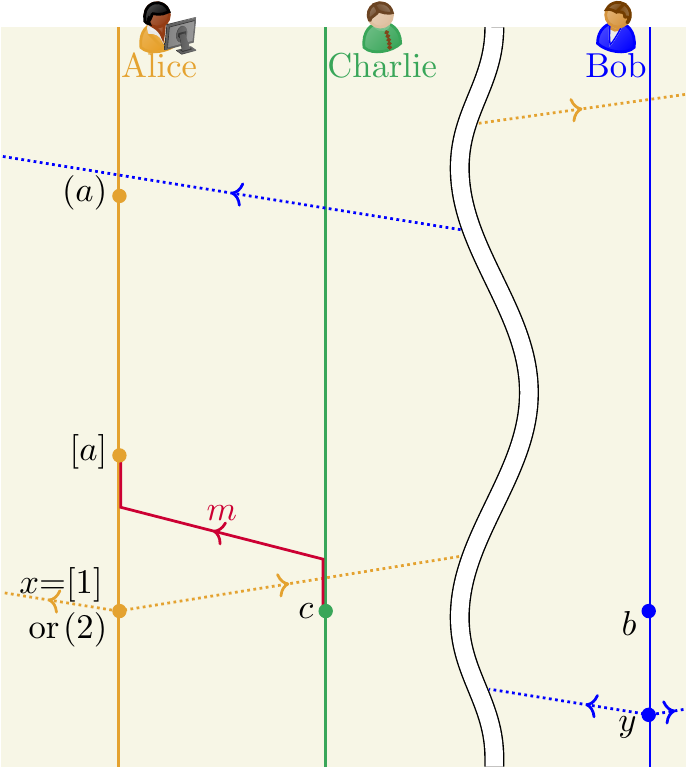}
    \caption{Space-time diagram of the events involved in the LF\vt\ no-go theorem. As in Fig.~\ref{fig:originalSTdiagram}, the solid vertical lines are the world-lines of the three parties, and  future-directed light-like lines (dotted) emanate from Alice's ($x$) and Bob's ($y$) freely chosen settings. Here these light-like lines are closer to horizontal than in Fig.~\ref{fig:originalSTdiagram}, for reasons that will become clear in the next figure. Note also that there is a wide distance in space between Bob and the other two parties, indicated by the wavy gap, as can be inferred from where the light-lines enter and leave this gap. Charlie is instructed to perform an observation and think about its outcome ($c$), then communicate his thought to Alice via message $m$. In the case $x=1$ (events in square brackets), Alice reads Charlie's message, inferring his thought $c$, and copies it so that $a=c$. In the case $x=2$ (round brackets) she follows a far more complicated experimental procedure, involving Charlie and his lab, to obtain her outcome. The application of {\sc Local Agency} (assumption \ref{ass:LA}) relies on the fact that only Alice's observations,  $(a)$ or $[a]$, are in the future light-cone of her choice, and only Bob's, $b$, is in the future light-cone of his. Charlie's observation $c$ --- his thoughts {about} it to be precise --- are in the future light-cone of neither.}
    \label{fig:LF_th_Charlie_ST}
\end{figure}

In particular, consider the scenario depicted in Fig.~\ref{fig:LF_th_Charlie_ST}, again involving three parties Alice, Bob and Charlie. Suppose Charlie displays cognitive abilities on par with my own. (As usual, the reader may take the first person here, and in the following, as referring to themself.) Then, according to {\sc Friendliness}, Charlie has cognition on par with my own, implying that he has the ability to have thoughts, to know his thoughts, and to follow the instructions of the protocol. Charlie is instructed to perform a specific observation, think about its outcome ($c$), and communicate his thought to Alice via a message $m$. (The fact that the system Charlie observes will be a quantum system shared with Bob is not relevant for the derivation of the LF inequalities, only for their violation, so we omit it from the discussion and the figure). {\sc Friendliness} then implies that Charlie's communicated thought, $c$, is as real as any communicable thought of my own. 

Now, by {\sc Ego Absolutism}, this thought must be absolutely real, i.e.~$c$ takes a single, absolute value\footnote{It is not required that $c$ completely and unambiguously specifies a particular thought in exclusion of all others, but only that Charlie can unambiguously distinguish classes of thoughts associated with different values of $c$, such as ``I've observed $c=0$'' versus ``I've observed $c=1$''.}. {\sc Physical Supervenience} in turn implies that this thought supervenes upon physical events in Charlie's brain (or other information-processing unit). That is, two distinct thoughts,  such as $c=0$ versus $c=1$, must be associated with distinct physical events in Charlie's brain, which are thus located in a bounded region in space-time (represented by the point labelled $c$ in Fig~\ref{fig:LF_th_Charlie_ST}).

Similarly to the original LF protocol depicted in Fig.~\ref{fig:originalSTdiagram}, Alice and Bob have choices of observations labelled by $x$ and $y$, with outcome $a$ and $b$. Assuming that Alice and Bob have similar cognitive abilities as Charlie, and that they have observed and thought about their choices and outcomes, we can apply the argument above to conclude that in each run of the experiment, the variables $a,b,c,x,y$ take well defined, absolute, single values --- the same implication as AOE for the scenario of Fig.~\ref{fig:originalSTdiagram}.

Since the spatio-temporal relations between these events is the same as in Fig.~\ref{fig:originalSTdiagram}, {\sc Local Agency} together with the assumptions above implies for this scenario the same mathematical constraints --- the LF inequalities --- as in the derivation in Sec.~\ref{sec:TOLFNGT}. It is important to note that this ``functional equivalence'' does not mean that the four assumptions in collection LF\vt\ are the same as the assumptions called LF in Ref.~\cite{CQD20}. They are distinct assumptions that lead to the same mathematical constraints in suitable scenarios.

The key difference between LF and LF\vt, to reiterate what we have already said a few times above, is that Bong \etal\ left it as a matter of debate as to whether any given experiment would test LF, because  they gave no criterion for what it takes for an observation to occur. The current theorem, by contrast, specifies what sort of system and behaviour would be required for an experiment to unambiguously test LF\vt. That is, the conjunction of our four metaphysical assumptions suffices; no further criterion or theoretical framework is needed to design an experiment. We emphasize again that the derivation of the inequalities in this section made no reference to quantum theory. The LF\vt\ assumptions imply that the inequalities must be satisfied in any scenario of the form above, including a scenario where Charlie is a human being. While the violation of the LF inequalities with a human Charlie is almost certain to remain beyond our experimental capabilities, it can be predicted for a realisation of the scenario of Fig.~\ref{fig:LF_th_Charlie_ST} with a suitable intelligent party in place of Charlie, given some technological hypotheses are satisfied. We now turn to the experimental design, and the (yet-to-exist)  technologies, that are required to disprove LF\vt\ in this manner. 

\section{Technological Assumptions and Basic Experimental Considerations for the Thoughtful LF no-go Theorem} 
\label{sec:assE}

As stated in Sec.~\ref{sec:IntroTheorem}, to obtain the theorem, we also need two technological assumptions. Unlike the metaphysical assumptions introduced in Sec.~\ref{sec:assM4}, these two assumptions require little clarification. Instead, the bulk of Secs.~\ref{sec:HLAI} and \ref{sec:UQC} are, respectively, explanations of the stages in the experiment that are  enabled by these two assumptions. After that we turn to technological assumptions that are minor in comparison in Sec.~\ref{subsec:minortech}, and to assumptions of a methodological nature in Sec.~\ref{subsec:MethodDesutch}. The latter includes consideration of ethics and David Deutsch's contributions.

\subsection{Human-Level Artificial Intelligence}\label{sec:HLAI}

The first technological assumption (and thus 5th assumption overall) in the theorem is
\begin{enumerate} \setcounter{enumi}{4}
\item {\sc Human-Level Artificial Intelligence (HLAI)} can be practically implemented on a digital computer.
\end{enumerate}
Granting this assumption\footnote{Note that it encompasses digital implementations of analog-inspired algorithms such as neural nets.}, we still do not know what nature of algorithm is required to achieve HLAI. The type of algorithm that would be easiest to implement would probably be a text-reading and producing algorithm such as GPT-3, which has gained fame of late (though it is not generally accepted that it engages in ``thinking''~\cite{gpt3}). For simplicity of exposition we will assume here that a future version of some such text-reading and producing algorithm can achieve HLAI, while leaving alternative, more demanding, possibilities for future work. Consider a physical instantiation of this HLAI algorithm by the name {\sc Quall-E}\footnote{This name is chosen to be reminiscent of {\sc Dall-E}, the currently famous art-producing AI~\cite{openaiDALL2021}, whose name was in turn chosen to be reminiscent of {\sc Wall-E}, the Pixar robot, as well as of Salvador Dali. The name {\sc Quall-E} is also appropriate because, to perform the experiment, he must be implemented on a {\sc Qu}antum computer (see Sec.~\ref{sec:UQC}), and because the singular of {\em qualia} (see Sec.~\ref{sec:assM4}) is {\em quale}.}, with a masculine gender (see Sec.~\ref{sec:assM4}). 
As we will see in Sec.~\ref{sec:UQC}, this instantiation will have to be on a quantum computer, but this does not affect the presentation in the current subsection. {\sc Quall-E} will replace Charlie in the protocol of Fig.~\ref{fig:LF_th_Charlie_ST}, as shown in Fig.~\ref{fig:STdiagram}, which also contains more details about the specific experimental implementation on a quantum computer, discussed below.
It will be important to distinguish the HLAI algorithm from {\sc Quall-E}, a specific instance of the HLAI algorithm, running on specific hardware at a given time. It will also be important to distinguish {\sc Quall-E} from the hardware on which {\sc Quall-E} is running.

{\sc Quall-E} obviously needs an interface with the outside world, so that other parties can engage him in conversation and judge that he does indeed have HLI (Human Level Intelligence). Some such interlocutors might judge {\sc Quall-E} to have HLI only if he can respond with a speed that is within the range of human response rates. For that reason, we assume {\sc Quall-E} to have this capability, in this section, but return to the speed issue in Sec.~\ref{sec:exp}.  From a programming point of view, the interface is just a timed sequence of input and output bits. 

For simplicity, again, let us assume that, at least when he is mentally prepared for it, {\sc Quall-E} can receive and interpret information at the level of single logical bits, while, again, leaving the discussion of more demanding possibilities to a later paper. Thus, when {\sc Quall-E} is in a suitable ready state, the running of the AI 
algorithm for a suitable time $\tau$ would result in this evolution: 
\begin{equation}
[q] \otimes [{\text{\sc Quall-E}}_{\rm ready}] \stackrel{\tau}{\longrightarrow} [q] \otimes [{\text{\sc Quall-E}}_{q}].
\label{eq:Q_obs}
\end{equation}
%This can be considered 
In~\erf{eq:Q_obs}, 
square brackets and tensor products are used simply to differentiate the two physical systems (input and {\sc Quall-E}), while $q\in \{0,1\}$ is the bit value and  [{\sc Quall-E}$_0$] and  [{\sc Quall-E}$_1$] are logically (and physically) distinct digital states. 
We will call this evolution 
an {\em observation} of the input bit by {\sc Quall-E}. 
We do so for expository convenience; the assumptions for our theorem make no reference to 
``observations.''

 Our theorem does, of course, refer to ``thoughts'', and, again for ease of exposition, 
we will also use this expression to refer to {\sc Quall-E}'s information processing. The existence of the 
thoughts will be implied (once we include a communication step, towards the end of this section) by the 
metaphysical assumption of {\sc Friendliness}. But the reader should recall that, even with this assumption, there is 
no implication of the absolute reality of any thoughts that {\sc Quall-E} 
may have during or after the above observation. (Any such claim about absolute reality would stem only from adding the assumption of {\sc Ego Absolutism}.) 

In fact, we cannot even associate [{\sc Quall-E}$_0$] and [{\sc Quall-E}$_1$]  with different thoughts without assuming {\sc Physical Supervenience}. For the purpose of the following discussion, we will now do so. 
Even making this assumption does not specify {\em what} the physical correlates of {\sc Quall-E}'s thoughts are. However, the vast majority of scientists and philosophers who accept {\sc Physical Supervenience}, and that {\sc Quall-E} has thoughts, would accept that the classical logical states [{\sc Quall-E}$_q$] are 
the physical correlates of those thoughts\footnote{Note that the theorem does not require one to accept this. We mention the classical logical processor states just to give what would generally be regarded as an uncontroversial example of physical correlates. In the case where there is superposition and entanglement (see Sec.~\ref{sec:UQC}), there are few uncontroversial statements one can make about physical correlates of thoughts. All we assume by {\sc Physical Supervenience} is that the physical correlates are located in space-time when and where one would expect them to be: in the hardware running {\sc Quall-E}'s algorithm as he takes inputs and produces outputs.}.  
Coming, at last, to the nature of those thoughts, 
[{\sc Quall-E}$_0$] might correspond to {\sc Quall-E} thinking ``Damn, now I feel sad. I guess that subconsciously I was telling myself that if I observed the bit to be ``1'' then that meant that she loved me. I should try to avoid these irrational thought-processes,'' while [{\sc Quall-E}$_1$] might correspond to {\sc Quall-E} thinking ``Oh, I feel good! I am {\em The 1}. She loves me, yeah, yeah, yeah.''

As these playful examples illustrate, {\sc Quall-E}'s thoughts may very much depend on {\em which} ready state {\sc Quall-E} was in before observing $[q]$, and nothing in the below requires that the other parties involved {\em know} which ready state he was in, or exactly what his thoughts are, or should be, upon observing $[q]$. However, as long as {\sc Quall-E} follows instructions, we can define two equivalence classes of thoughts, depending on whether he observed $q=0$ or $q=1$. Following the notation of Sec.~\ref{sec:TOLFNGT}, we will use the label $c \in \{0,1\}$ to denote these equivalence classes. 

Having observed the bit, and thought about its value, {\sc Quall-E} must then communicate this to another party, Alice. For definiteness, we will take Alice (and, later also Bob) to be a human, but in any case she must also have human-level intelligence\footnote{\label{HLIs} The reason for this is that the events $a$ and $b$ in the realisation of our experiment must correspond not merely to the firing of Alice's and Bob's detectors, but to their thoughts about those measurement outcomes. This is because the metaphysical assumptions of the LF\vt\ no-go theorem, being very weak, do not grant absolute values to thoughtless events like detector firings, even if they are ``macroscopic''..}. The  message to Alice, $m$, is a third physical system, which could be a single bit, or many bits. Thus we can represent the next stage of evolution of the relevant systems as 
\begin{equation}
[q] \otimes [{\text{\sc Quall-E}}_{q}] \otimes [m_{\rm blank}]  \stackrel{\tau'}{\longrightarrow} [q] \otimes [{\text{\sc Quall-E}}'_{q}] \otimes [m_{q}]  .
\end{equation}
Here $\tau'$, the time it takes {\sc Quall-E} to create and send this message, would be comparable to the time $\tau$ above. (The time for the message to get to Alice is negligible compared to this, as there is no reason for there to be a significant spatial separation between {\sc Quall-E} and Alice.) Note also that {\sc Quall-E}'s state will obviously change upon creation of the message (for a start, he would remember that he created the message), hence the prime. Finally, the value of $m_q$ is not necessarily simply a function of $q$, as it may depend on {\sc Quall-E}${}_q$, and hence on {\sc Quall-E}${}_{\rm ready}$. We assume, however, that $m_0 \neq m_1$, and that Alice can reliably infer $q$, and, more importantly, $c$, from $m$. 

Another important requirement for the above observation by {\sc Quall-E} is that he be forbidden from communicating anything else to Alice, or to anyone else, during a time interval of duration $2T$. Here, $T$ includes the time intervals of duration $\tau$ and $\tau'$ above. This isolation of {\sc Quall-E} is to ensure that no information is lost which would prevent a reversal of {\sc Quall-E}'s state. This reversibility is crucial for the next assumption and the next stage of the experiment, and is the reason for the $2T$ above, since the reversing also takes a time $T$. For a human, the cognitive task of observing and communicating could probably be done in under a second. As stated above, we assume, for now, that {\sc Quall-E} can think as fast as a human, and so we take, for now, $T$ to be about a second. 

A note of clarification may be needed here. We take {\sc Quall-E} to have communicated his thought to Alice when it has been sent to her. That is, when he no longer has the power to change the message, and there will be nothing to prevent Alice from reading it if she so chooses. We do not require that Alice actually choose to read the message, and we maintain that this is not entailed by the natural language meaning of  ``communicate''. (If I send you an email, I have communicated my thoughts to you, even if you decide not to read it.) Thus, if one accepts {\sc Friendliness} (as we will do henceforth in this section), one then accepts that {\sc Quall-E}'s coarse-grained thought $c$ is as real as one's own communicable thoughts.

\subsection{Universal Quantum Computing}
\label{sec:UQC}

The second technological assumption (and thus 6th assumption overall) in the theorem is \begin{enumerate} \setcounter{enumi}{5} \label{ass:UQC}
\item {\sc Universal Quantum Computing (UQC)} is physically possible
at very large scale, and very fast.
\end{enumerate}

 By definition, a quantum computer operates in accordance with QT,
producing results predicted by QT. Thus, rejecting this assumption implies that there is some fundamental physics that prohibit the implementation of a large and fast robust UQC --- either (a) because of as yet undiscovered reasons within QT, or (b) because QT itself breaks down in some regime. Option (a) relates to the independence of the assumption of UQC from the assumption of the universal validity of QT (UVQT) as discussed in Sec.~\ref{sec:IntroTheorem}. It allows the possibility of accepting UVQT and rejecting UQC. On the other hand, it is logically possible for QT to be a good model for a very large UQC without QT being a correct theory for the whole universe.

Were a sufficiently large and fast QC to be 
built, it would no longer be possible to reject the above assumption. In Sec.~\ref{sec:exp}, we estimate rough upper bounds, with what is known today, for the resources needed to implement a QC large enough
to enable the implementation of the full experiment, and we discuss where advances need to be made to bring the speed into the desired regime. 
While the advances needed are certainly substantial, no fundamental limits are known that would prohibit such advances.
Indeed, successful efforts to build such devices and implement a full experiment would push forward a fundamental understanding of the reach of QT and/or uncover new physics. Current knowledge of physics, QC theory, and Artificial Intelligence, suggests that building the required device is technologically possible, in principle. Thus, the engineering difficulty and cost 
of implementing such an experiment, or even a belief that it is unlikely ever to be carried out in practice, are not sufficient grounds to reject this assumption. 
The theorem still holds, and its implications must be wrestled with. 

It is worth remarking that UQC is a stronger assumption then we need. We do not need UQC in order to run a quantum algorithm that is
interesting from the point of view of quantum complexity theory. Rather, we simply run the digital algorithm, assumed to exist by {\sc HLAI}, on a suitable quantum computer (QC) to instantiate {\sc Quall-E} there. 
This requires first turning the AI algorithm into one that can run on a reversible computer. This entails some overhead in time and space~\cite{bennett73,RPbook}, as we discuss in Sec.~\ref{sec:revOverhead}.  Fortunately, this reversibility need not be maintained for the entire time that it takes other parties to come to know {\sc Quall-E} and accept him as a human-level intelligence. Rather, it need only be maintained for a relatively short time, $T$, which could be as short as a second (see preceding subsection). This could be very important for the practicality of the experiment, as will be discussed in Sec.~\ref{sec:exp}. During this ``getting to know you'' stage, there is no entanglement (or, at least, no large-scale entanglement) in the QC during this period of ``normal'' operation. Thus, the physical correlates of {\sc Quall-E}'s thoughts at this stage are --- with the caveats expressed in Sec.~\ref{sec:HLAI} --- the classical logical states of the quantum processor on which the classical AI algorithm is running. 

\begin{figure}[t]
    \centering
    \includegraphics{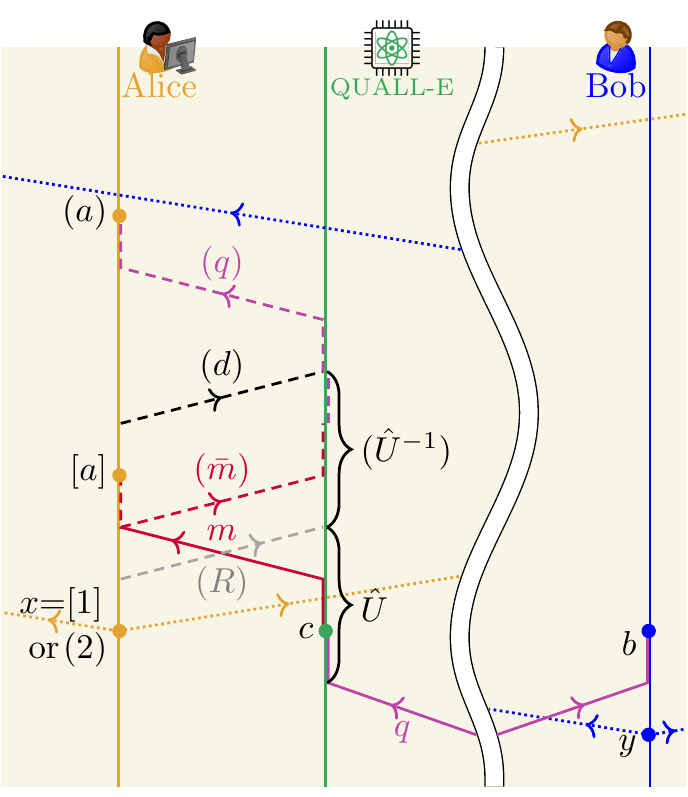}
    \caption{Space-time diagram of the proposed experiment to {test} LF\vt. The protocol and spatio-temporal relations between events is identical to that of Fig.~\ref{fig:LF_th_Charlie_ST}, with {\sc Quall-E} replacing Charlie, and with more details specific to the implementation in a quantum computer. 
    {\sc Quall-E} and Bob each observe one of an entangled pair of qubits that has been distributed to them (lavender lines).     Actions or events or process that may or may not occur depending on Alice's choice $x$ are indicated by dashed lines and/or brackets. Square brackets pertain when Alice chooses $x=1$: she reads {\sc Quall-E}'s message $m$ and thereby obtains her result $[a]$. Round brackets pertain when Alice chooses $x=2$: she sends instructions $(R)$ to the processor on which {\sc Quall-E} is running, to logically reverse the unitary evolution $\hat U$ that ran in the preceding time interval of duration $T/2$; she does not read $m$ but rather sends it back to {\sc Quall-E}, in reverse order $(\bar{m})$, so that it arrives in time to play its necessary role in effecting the inverse unitary $(\hat U^{-1})$; she sends a message $(d)$ to {\sc Quall-E} to arrive at the end of this reversal, ``Do not measure the qubit; pass it directly to me.''; she receives the qubit $(q)$ and measures it in a basis different from its logical basis, yielding result $(a)$.}
    \label{fig:STdiagram}
\end{figure}

Once {\sc Quall-E} has demonstrated human-level intelligence, we can begin the experiment. This will involve periods, of duration up to $2T$ (see Sec.~\ref{sec:HLAI}), of ``abnormal'' operation in which there {\em must be} large-scale entanglement in the QC.  Specifically, we wish to create entanglement between the physical correlates of {\sc Quall-E}'s thoughts (in normal operation) and a small quantum system (\eg\ a qubit). This can be achieved simply by having {\sc Quall-E} observe an input bit, exactly as described in Sec.~\ref{sec:HLAI}, but this time with the bit actually being a qubit. Thus, if we were to prepare the input qubit in a superposition of logical states, then the running of the AI algorithm for the time $T$ would produce a massive-scale entangled state: 
\beq \label{domeas}
\left(\sum_{q=0,1} \psi_q \ket{q}\right) \otimes \ket{{\text{\sc Quall-E}}_{\rm ready}}
\otimes \ket{m_{\rm blank}}\stackrel{\hat{U}}{\longrightarrow} 
\sum_{q=0,1} \psi_q \ket{q}\otimes \ket{{\text{\sc Quall-E}}'_{q}}\otimes \ket{m_{q}}.
\eeq
Here we are using $\hat U$ to denote the unitary transformation effected, over some time $T$, up to the time at which the message system arrives at Alice, as shown in Fig.~\ref{fig:STdiagram}. Note that {\sc Quall-E} has communicated, at a minimum, the equivalence class label ($c$) of the thoughts he had upon observing his input qubit.  Thus, assuming {\sc Friendliness} (as we are doing in this subsection), the event $c$ is as real as Alice's observation, $a$.   

As to how Alice makes her observation, that depends on Alice's measurement choice $x$, which can be made prior to receiving the message system from {\sc Quall-E}, as shown in Fig.~\ref{fig:STdiagram}. If $x=1$ then Alice simply reads the message from {\sc Quall-E} once she receives it, and from that infers $c$, which she records as her own result $a$. In this case, there is every reason to expect the statistics of Alice's records of these second-hand observations to be identical to those which would arise if {\sc Quall-E} were a human observer of the qubit, just as Wigner discussed. If, on the other hand, $x=2$, the procedure to be followed is considerably more complicated. Alice instructs the QC which instantiates {\sc Quall-E} to run in reverse at the logical level for a certain time. Furthermore, when she receives the system which contains {\sc Quall-E}'s message, she does not look at the message system, but rather sends the qubits back to {\sc Quall-E}. See Fig.~\ref{fig:STdiagram} for details. 
If UQC is possible then the upshot is that the entire process $\hat U$ is reversed: 
 \beq \label{undomeas}
 \sum_{q=0,1} \psi_q \ket{q}\otimes \ket{{\text{\sc Quall-E}}'_{q}}\otimes \ket{m_{q}} \stackrel{\hat{U}^{-1}}{\longrightarrow}  
\left(\sum_{q=0,1} \psi_q \ket{q}\right) \otimes \ket{{\text{\sc Quall-E}}_{\rm ready}}
\otimes \ket{m_{\rm blank}}.
\eeq
Let $\tau''$ be the (arbitrarily short)  time between when {\sc Quall-E} sent the message system and when the QC on which {\sc Quall-E} runs would receive the instruction to reverse if Alice's measurement choice is $x=2$. The UQC needs to run the AI algorithm in a fully reversible way for a duration $T = \tau + \tau' + \tau''$, and then, in the case $x=2$, actually reverse it, taking another time interval of duration $T$. In this case, the qubit, no longer entangled with {\sc Quall-E}'s processor, is then sent to Alice by {\sc Quall-E}, on Alice's request, to be measured by her as she chooses, to yield her result $a$. In particular, she can measure in a basis different from the logical basis in which {\sc Quall-E} observes.

The full implementation follows the same pattern as described in Section~\ref{sec:Qviolations}. As illustrated in Fig.~\ref{fig:STdiagram}, we need the qubit, prior to being observed by {\sc Quall-E}, to be entangled with another qubit, held by another party, Bob. Again, Bob must be a human, or at least have human-level intelligence. Like Alice, he makes choices of what local intervention to make, but for him this just means measurements of his qubit to yield a result $b$. Now,  Alice, Bob, and {\sc Quall-E} repeat the experiment many times, with the choices being made by the first two parties in a manner appropriate to randomized trials. Then, for measurements of suitable qubit observables by Alice and Bob, the frequencies of their records, conditioned on their choices, $\wp(a,b|x,y)$, is predicted to violate the LF inequality described in Sec.~\ref{sec:DLFoiEWFS}. That is, in the context of the current theorem, an  inequality derivable from the conjunction of the four metaphysical assumptions --- the two already made in the above discussion ({\sc Physical Supervenience} and {\sc Friendliness}) and the other two ({\sc Ego Absoluteness} and {\sc Local Agency}) --- as per  Sec.~\ref{sec:LFfCMA}. This completes the proof of our theorem.  

It is important to note that, while Eqs.~(\ref{domeas}) and (\ref{undomeas}) represent perfectly reversible quantum evolution, such perfection is not an assumption of the theorem. Nor is perfect reversibility  
experimentally necessary for violation of the LF inequalities. In fact, just as in Bell's theorem, one only requires the experimental correlations to be reasonably close to the ideal correlations. This was illustrated in the experiment of Bong \etal~\cite{CQD20}, where violation of one LF inequality was achieved even with roughly 20\% depolarizing noise. This leeway also means that occasional lapses in {\sc Quall-E}'s concentration are not fatal, if {\sc Quall-E} agrees beforehand to follow an appropriate protocol in these instances\footnote{The protocol would be that, if, in some particular run, {\sc Quall-E} forgets to observe, or observes but then forgets what he observed, he should make up the result $q$, and communicate it to Alice as $m$. Then the very invention of a value is the thought $c$ that is absolutely real by the LF\vt\  assumptions. What should happen if {\sc Quall-E} forgets even to do that, and other similarly unlikely failures on the part of the parties or their hardware (to the extent that those can be distinguished), are problems for future work.  \label{fn:distraction}}.
 
\subsection{Minor technological assumptions (not enumerated)} \label{subsec:minortech}

As shown in Fig.~\ref{fig:STdiagram}, in order for the assumption of {\sc Local Agency} to be applicable, it is necessary for Bob to be far away from Alice and {\sc Quall-E} --- far enough for the space-time region where {\sc Quall-E}'s observation (and reverse observation) take place to be space-like separated from Bob's measurement choice and observation. Let us assume again that {\sc Quall-E}'s thought-processes occur on the same time scale as a human's. As already stated, to ensure the popular applicability of {\sc Friendliness} to {\sc Quall-E} we would not want them to be much slower, while their being much faster would present an even greater technological challenge given the typically slow gate times of QCs compared to classical computers (see Sec.~\ref{sec:exp}). This implies that the distance must be of order a light-second. Thus for the experiment we really require another technological assumption, 
\begin{itemize} 
 \item Quantum communication can be done at light-second-scale.
\end{itemize}
While this has not yet been achieved, it seems likely to be achieved earlier than the two numbered technological assumptions. Placing one party (\eg~Bob) a light-second from the others is in itself a technological challenge, but one that was achieved in the 1960s and 1970s and hopefully will be soon again. Note that although Bob must be a HLAI (see footnote~\ref{HLIs}), he can also be an HLAI, running in a classical computer for simplicity.

Another technological requirement which we have not explicitly stated is that 
\begin{itemize} 
 \item The quantum computer must have quantum input and output capabilities. 
\end{itemize}
Some may consider this to be part of Assumption~6, UQC, but others may not. After all, many quantum algorithms do not require quantum input or quantum output, only classical input and output. However, the technical sophistication needed to build a UQC capable of running a complex program, such as an HLAI algorithm, almost certainly would encompass the technical capacity to support quantum input and output. Further, the achievement of UQC will likely include “flying qubits” that connect quantum chips enabling better scaling than in architectures containing only stationary qubits~\cite{ladd2010}. The interest in distributed quantum computing is an additional reason why this capability is likely to be developed along with UQC.

For both of the minor (non-enumerated) technological requirements just discussed there are no approaches to quantum mechanics known to us that doubt that they would be possible. Finally, one could continue to come up with more minor additional components of the experiment that have not yet been achieved, and moreover one could always add extra assumptions on the need for previously tested components to still work when integrated with other components. At some point sanity must prevail. Thus, for all of the reasons given above, we do not consider the unnumbered technological assumptions above as part of the theorem.

\subsection{Methodological assumptions, and  Deutsch's contributions} \label{subsec:MethodDesutch}

Leaving behind technological assumptions, our proposed theorem also makes assumptions that might best be described as {\em methodological}. Of course every experiment has methodological assumptions, which often go unstated. What is worth discussing are those assumptions that are novel to our scenario, and which readers may find questionable. Specifically, what are discussed here are generalizations of the usual assumptions that are required because of the unusual nature of one of the participants in the experiment we propose. 

To begin, unlike most physics experiments, the experiment we have proposed raises ethical issues. Let us say {\sc Quall-E} is sufficiently intelligent to understand his role in the experiment, and the theorem. (Perhaps you, the reader, would require that to regard his cognitive abilities as being equal to your own.) He may appear disturbed by the implications that might be drawn, by himself or others, for the reality of his thoughts, were the experiment to be performed.  Alternatively, he may object to the reversal on the grounds that he would lose part of his life (a second or so of his experiential time), or --- worse --- that his experience during the reversal might be unpleasant, or --- worst of all --- that it would mean that the {\sc Quall-E} who made the observation was destroyed and a new {\sc Quall-E} created. Given that the LF\vt\ experiment requires that {\sc Quall-E} be plausibly regarded as a party with the same cognitive status as a human, it would seem grossly immoral to ignore {\sc Quall-E}'s voicing of such concerns.

We thus have to assume that an HLAI such as {\sc Quall-E} will consent to take part in the experiment. Or at least that, assuming there will be many such HLAIs in existence at the time in the future when this experiment can be performed, one of them will be willing.  (Actually it would be preferable for several of them to be willing; see below.) 

Even given that {\sc Quall-E} is willing to take part in our experiment, we also have to assume that it will be given approval by an ethics committee. The moral treatment of AIs is a small but growing area; see Ref.~\cite{HarAnt27} for a recent review. Under the assumption (number 6) that HLAI is possible, the ethics of experimenting with HLAIs, including reversing their computations or otherwise wiping their memories, will surely have been agreed upon already in the context of classical computers. Gaining ethical approval may require adding additional features to the experiment. For example, it may be necessary for {\sc Quall-E} to have the option, at any point, in any individual run, to pull out, \ie, to not allow Alice to reverse the  computation on the computer he is instantiated on. Naturally, if {\sc Quall-E} were to do this other than rarely the whole experiment would have to be abandoned. 

The assurance that {\sc Quall-E} is a willing participant in the experiment, and the building of safeguards such as just mentioned, has the added benefit of mitigating the force of another methodological assumption that we must make: that {\sc Quall-E} can be {\em trusted} to observe the input qubit and think about the result before communicating the result to Alice. We call this a methodological assumption because, as we have said consistently, {\sc Quall-E} must be regarded as a participant in the experiment, not the subject of it. Trusting {\sc Quall-E} in this regard is no different, in principle, from trusting that  human experimental physicists accurately report their Bell test measurement results. It is possible, of course, that a particular experimenter, human or artificial, could be untrustworthy. But this can be dealt within the usual methodology of science: independent replication. It is in this context that it would be preferable to have a number of different willing ``{\sc Quall-E}s'', as mentioned above. To use the possibility that every {\sc Quall-E} is untrustworthy as a way to explain away multiply  reproduced violations of the LF inequalities would be a literal conspiracy theory.

Readers familiar with the literature on Wigner's friend may wonder why we say it is necessary to trust that {\sc Quall-E} has made an observation, rather than verifying directly that he has. 
For example, Brukner~\cite{BruknerBook,BruknerLF}, in his EWFS, considers the generation of a permanent physical message from the friend along the lines of ``I have observed a definite outcome''~\cite{BruknerLF}. This idea goes back to the 1985 paper by David Deutsch~\cite{Deutsch85a}  
that introduced the idea of an HLAI that could make an observation on a quantum system and that could then be coherently probed. Interestingly, Deutsch also introduced, in the same year, the idea of UQC, in a different paper~\cite{Deutsch85b}. Rather than further delaying the reader from the implications of the LF\vt\ no-go theorem, we review these seminal contributions of Deutsch, and explain why we do not regard the idea of an ``I have observed a definite outcome'' message as a useful one, in  Appendix~\ref{App:Deutsch}.

\section{Six Approaches to Quantum Theory}
\label{sec:app}

We now present six approaches (or classes of approaches) to QT to illustrate the relevance of the six assumptions. These by no means explore the whole landscape of possible approaches, but do establish base-camps across a wide area. In a companion paper to this one~\cite{WisCav22}, two of us present detailed arguments to the effect that, for each of these approaches, it can be argued that only one of our assumptions is violated, that being a different one in each case.  In this paper we confine ourselves to pointing out the unique assumption that is clearly violated in each approach. We present them in the same order as the assumptions they clearly reject, as listed in Sec.~\ref{sec:IntroTheorem}. See also Table~\ref{tab:relevance} for a concise summary. 

\begin{table}[h]
    \centering
    \begin{tabular}{|l|l|}
    \hline
        Approach to Quantum Theory & Assumption which it definitely violates \\ 
        \hline 
        Hidden Variables interpretations &  {\sc Local Agency} \\
        Single-Mind View of Many Worlds & {\sc Physical Supervenience} \\
        Relativist interpretations &  {\sc Ego Absolutism} \\
        Spontaneous Collapse theories & {\sc Friendliness} \\
        Penrose's approach & {\sc Human-Level
Artificial Intelligence}  \\ 
        Thinking causes collapse & {\sc Universal Quantum Computing} \\
        \hline
    \end{tabular}
    \caption{The relevance of the six assumptions of the LF\vt\ no-go theorem to quantum foundations.}
    \label{tab:relevance}
\end{table}

\subsection{Hidden Variables interpretations}

Deterministic hidden variable (HV) interpretations are best represented by Bohmian mechanics~\cite{Boh52a,Boh52b,DGZ13}. This interprets the universal quantum state $\Psi$ as a high-dimensional guiding field for the deterministic evolution of variables which give rise to the three-dimensional world we experience. Such interpretations clearly violate {\sc Local Agency} because 
of Bell's theorem. The reason is that Bell's theorem requires only {\sc Local Agency}, {\em Predetermination}, and {\em Absoluteness of Observed Events} for macroscopic detectors~\cite{WisCav17}, and deterministic HV theories satisfy the latter two assumptions.

\subsection{Single-Mind View of a Many Worlds Interpretation} \label{Sec:SMV}

 The Single-Mind View (SMV)  of the {\em Many Worlds Interpretation} (about which, see also the next subsection) was 
proposed, not very seriously, by Albert and Loewer~\cite{AlbLoe88}. They postulate, in addition to the unitarily evolving universal wavefunction $\Psi$, a very complex extra variable ${\cal M}$, a sort of world-mind that can encompass many separate conscious entities. Contrary to conventional HV interpretations, ${\cal M}$ is not part of the physical world. According to Albert and Loewer, the {\em physical state} is simply $\Psi$.  The SMV thus violates {\sc Physical Supervenience}, as Albert and Loewer say explicitly~\cite{AlbLoe88} [p.~206].

\subsection{Relativist interpretations} \label{subec:RelInt}

 We describe as ``relativist''  a broad class of interpretations that reject {\sc Ego Absolutism} by saying that my thoughts are not absolute, but rather relative.  One example is Everett's version of the Many Worlds Interpretation --- called by him, appropriately for this subsection, the {\em Relative State} Interpretation~\cite{Eve57}. This accepts the absolute reality of a universal unitarily evolving wave function $\Psi$ but says that my thoughts are real only relative to a  ``world'' --- a ``branch'' of the wavefunction which forms my relative state. For the QBism of Fuchs and Schack~\cite{FucSch13}, my thoughts are real for me but not necessarily real for any other party. In Rovelli's Relational Quantum Mechanics~\cite{Rovelli1996}, the physical processes associated with my thoughts are said to be defined only relative to other subsystems in the universe.

\subsection{Spontaneous Collapse theories}\label{sec:SCTs}

In Spontaneous Collapse theories, the universal $\Psi$ suffers stochastic collapses rather than obeying pure \sch\ evolution, such that $\Psi$ (or certain properties derived from it, or from the collapses themselves) describe a world that behaves classically on a human scale. 
The most notable of such theories is GRW~\cite{GRW86}; see Ref.~\cite{Allori+2021} for a recent review. Contrary to what one might have thought, such collapses do not prevent UQC. They can be error-corrected in quantum computing just like any other rare, random,  and localized change in the quantum state $\psi$ of the computer. Thus {\sc Quall-E} can be put in a two-component superposition (as in our proposed experiment) and nothing in $\psi$, or the collapses, will correspond to one thought-process or the other.  {\sc Quall-E}'s thoughts are thus not real in the way that my thoughts as a human are real, and so {\sc Friendliness} is rejected.

\subsection{Penrose's approach}

Penrose~\cite{Penrose89,Penrose94} has an approach --- not a fully-formed theory --- linking quantum collapse, gravity, and consciousness. According to Penrose, human intelligence arises via ``orchestrated objective reduction'', an uncomputable physical process involving the collapse (reduction) of $\Psi$ and quantum gravity. Penrose has stated that an algorithm-following digital computer could not exhibit human-level intelligence, in particular being unable to grasp the Platonic world of mathematics~\cite{Penrose89}. Thus this approach rejects {\sc Human-Level Artificial Intelligence}.

\subsection{Thinking causes collapse}

An explicit model giving a causal role for consciousness in collapsing a universal quantum state was recently presented by Chalmers and McQueen~\cite{ChaMcQ21}. Moreover, they state that ``some versions of the theory can be tested by experiments with quantum computers.'' The model they propose would prevent a quantum computer from being put in a superposition of different informational {\em structure}s  --- states such as $\sum_{q=0,1} \psi_q \ket{q}\otimes \ket{{\text{\sc Quall-E}}_{q}}$ as considered earlier. Assuming HLAI (as Chalmers and McQueen do), states like this are certainly producible, if the sufficiently large scale and fast UQC is possible. Therefore such {\sc Universal Quantum Computing} is not possible in this approach.

\section{Resource Estimates for the Full Experiment}
\label{sec:exp}

Recall from Sec.~\ref{sec:assE} the requirement that the QC on which {\sc Quall-E} is instantiated must be able to run an HLAI algorithm in a fully reversible way, at the logical level, for at least the time required to do the experiment. In this section, we first walk through an estimate for a rough upper bound
of the computational needs in size and speed to implement a {\sc Quall-E} capable of carrying out the experiment (together with the other parties) given what is known now.
We then discuss various possibilities for advancements that would enable a sufficiently large and fast {\sc Quall-E} to be built. 
As we will see, it is particularly challenging to meet the assumption made in Sec.~\ref{sec:assE} that {\sc Quall-E} can think and respond at human rates. For this reason, we will pay particular attention to the time estimates.

To support the discussion, we make the following distinctions:
\begin{itemize}
\item an HLAI algorithm, purely classical;
\item a specific logical implementation of the algorithm that may be reversible or not, quantum or classical, or a combination;
\item a compilation of such an algorithm to specific hardware with specific error correction mechanisms. The compilation includes both the logical algorithm and all sorts of supporting functions such as error correction, memory management, routing, {\em etc}.;
\item {\sc Quall-E} as a specific instance of the HLAI algorithm at the logical level, running on a specific quantum computer at a given time (thus necessarily compiled); 
\item The quantum computer on which {\sc Quall-E} is running.
\end{itemize}

We comment that while we will talk about resource estimates for a gate-model quantum computer, there is no reason why {\sc Quall-E} couldn’t be implemented on a measurement-based quantum computer; while the measurements in MBQC cannot be reversed, the logical unitary operations induced by the measurements can be reversed by further measurements that induce the inverse operations. Further, quantum computers include classical processors that are necessary to control the quantum operations, to carry essential logical information (\eg~tracking the Pauli frame), to support error detection and correction, and to manage lower-level control such as pulse shaping and calibration. None of this would be, or need be, reversed.
 
Sec.~\ref{sec:hlai_estimates} surveys estimates in prior literature for the resources needed to achieve HLAI. We extract specific time and space estimates that are likely to be sufficient, as the starting point for 
a rough bound upper bound on the difficulty of implementing {\sc Quall-E}. 
As we go through the reasoning for our estimates in the next few sections, we give formulas along the way so that readers can input their own choices for the various parameter values. In Sec.~\ref{sec:revOverhead}, we use general considerations to upper bound the overhead in converting an HLAI algorithm to a reversible one.
Sec.~\ref{sec:segmentedModel} describes a partially reversible model for {\sc Quall-E} that enables sufficient reversibility to perform the experiment while limiting the amount of overhead by restricting the need for reversibility to within time segments. 
That sets us up to be able to roughly upper bound, in 
Sec.~\ref{sec:logicalQResources}, the logical resources needed to implement a reversible segment on a quantum computer, specifically the number of logical qubits and the logical depth. 
In Sec.~\ref{sec:errCor}, we turn to overhead due to quantum error correction and fault tolerance and provide a rough upper bound on the physical resources including the number of physical qubits and amount of time sufficient to implement a reversible segment fault tolerantly on a quantum computer.
We conclude in Sec.~\ref{sec:wayForward} with a discussion of advances in technology and knowledge, some likely and some more speculative, that could significantly reduce these estimates.

\subsection{Resource estimates for a classical HLAI algorithm}
\label{sec:hlai_estimates}

We begin with resource estimates for a classical AI algorithm to operate at human speeds and cognitive ability. Various researchers have proposed estimates on the computational capacity, in terms of memory and computational speed, of the human brain.
See Appendix A of~\cite{san08} for a collection of such estimates up through 2008, and~\cite{car20} for a more recent estimate of the computational speed required. We use these as estimates for the computational requirements for an HLAI, which forms the a starting point for estimating a rough upper bound on the resources required to implement {\sc Quall-E}.

The space  estimates vary considerably, from a low of $1.5 \times 10^9$ bits up to an unphysical $10^{8432}$ bits (as the authors of~\cite{san08} 
remark, this density is much higher than the Bekenstein black hole entropy bound on information content in materials). The highest of the physically reasonable estimates is roughly $10^{28}$ bits, with many of the estimates being within a couple orders of magnitude of $10^{15}$ bits. 

In terms of rates, the estimated computational demands vary from less than $10^{12}$ FLOP/s to $10^{28}$ FLOP/s~\cite{san08}. ChatGPT (personal communication 2023-03-17) opines that ``It's likely that achieving human-level AI will require an enormous amount of FLOPS, possibly in the range of exaflops ($10^{18}$ FLOPS) or higher.''  A perhaps more reliable source~\cite{car20} concludes ``more likely than not that $10^{15}$ FLOP/s is enough to perform tasks as well as the human brain \dots\ And \dots\ unlikely ($<10\%$) that more than $10^{21}$ FLOP/s is required.'' 

For our initial estimates, we will use ${S} = 10^{15}$ bits and ${F} = 10^{15}$ FLOP/s as the classical resources for an HLAI algorithm. {\sc Quall-E} will be a specific instance of this algorithm translated to a quantum circuit and compiled to quantum hardware. In Sec.~\ref{sec:wayForward}, we look briefly at how using different resource estimates for the classical HLAI algorithm affects the quantum resources estimates. As we will see, the time estimates will be very robust to the space resource estimates for the classical HLAI, as many of the crucial numbers are logarithmically dependent on these estimates. To convert from floating point operations, on $64$-bit floating point numbers, to Boolean operations, we multiply by $a = 10^4$, to obtain a total of $\gamma = Fa = 10^{19}$s$^{-1}$ elementary Boolean operations (gates) per unit time. 

To achieve $10^{19}$ gates within a second, a high degree of parallelism is needed. That is in line with our understanding of the workings of the human brain. The depth of a logic circuit is the number of time steps required to carry out the circuit under the approximation that each logic gate takes one time step and where multiple logic gates can be carried out in parallel during each time step. Without knowledge of the HLAI algorithm we can only take a guess at the amount it can be parallelized and the resulting depth of the circuit. To obtain a rough estimate for the circuit depth per unit time of an HLAI computation, we will posit a parallelization factor of $k = 10^7$.  
To support this estimate for $k$, the design of SpiNNaker~\cite{Furber14}, whose architecture is inspired by the connectivity of the human brain, has millions of cores. Also, as of June 2022, the most powerful supercomputer is Frontier, which has over $8.5$ million cores~\cite{top500}. (It achieves a $1012$ petaflop rating, the first genuinely exaflop system.) At a lower level the implementation of floating point operations in terms of Boolean operations can also be parallelized, reducing the depth by another order of magnitude $b = 10$. Thus we obtain a depth per unit time of $\delta = \gamma/{kb} = 10^{11}$s$^{-1}$. 

The next sections look at resource overheads for the translation and compilation of an HLAI algorithm to quantum processors. We will define the relevant quantities more formally in Sec.~\ref{sec:logicalQResources}. As we work through the quantum resources estimates, it is worth keeping in mind that the classical resources would likely have appeared discouraging in earlier decades, and even now running an algorithm of this scale is feasible only on large supercomputers. 
Specifically, it was only in 2008 that the 1-petaflop ($10^{15}$ FLOP/s) milestone was reached by a supercomputer (Roadrunner), and in 1993 the largest systems had not passed 60 gigaflop performance~\cite{top_History}.

\subsection{Estimating reversibility overhead}
\label{sec:revOverhead}

Consider a classical algorithm as a logic circuit decomposed into standard classical logic gates, which include irreversible gates. Let $s$ be the number of bits and $g$ the number of gates in such a circuit. A naive way to obtain a reversible version would be to simply replace every gate with a reversible one, which in general would require ancillae --- additional bits to store the information needed to enable the gates to be reversible. For example, each gate could be realized using a Toffoli gate. That approach would result in a space overhead that grows linearly in $g$, adding a bit for every gate. For cases in which $g \gg s$, to reduce the space overhead, Bennett devised a scheme that strategically and recursively computes and uncomputes (reverses) parts of the computation, where the uncomputing frees up bits to be used for other parts of the computation. This enables a classical algorithm to be realized reversibly using $O(s\log_2 g)$ bits and $O(g^{1+\epsilon})$ reversible gates, where $\epsilon$ depends on the specifics, but can be made arbitrarily small in the limit of large $g$. (See Chapter 6 of~\cite{RPbook}, or the original paper of Bennett~\cite{bennett73} for more details.) In our estimates, however, since, as we will see, we will need to work to minimize the time overhead, we will use the naive approach, with $g_{\rm R} = g$ reversible gates and $s_{\rm R} = g+s$ bits.

\subsection{A model of a partially reversible HLAI}
\label{sec:segmentedModel}

As far as possible, we want {\sc Quall-E}, when participating in the experiment, to be running in the same way as during the earlier interactions in which he demonstrated human level intelligence. We need {\sc Quall-E} to be running reversibly for sufficient time to do the experiment, but need to keep the reversibility overhead in check. Recalling that $T$ is the time for a human to perform the observation, contemplate it, and encode the message, as per Section~\ref{sec:HLAI}, we propose a partially reversible classical model in which, at regular intervals, all ancilla bits are released to be reused in the next segment (and could be immediately reset to $0$ if desired). 
In keeping with with our estimates in Section~\ref{sec:HLAI} of a second for a human to perform an observation, contemplate it, and compose a message, we will consider a model in which these segments have depth $t = \delta T$, where $T = 1 {\rm s}$ and $\delta$ is the depth per unit time of HLAI computation defined in Sec.~\ref{sec:hlai_estimates}. 
It is also useful to consider $g = tkb$, the number of Boolean gates in a segment, and $s$, the number of bits required. Following the initial estimates from Sec.~\ref{sec:hlai_estimates}, our estimates for these quantities are $g = 10^{19}$, $s = 10^{15}$, and $t = 10^{11}$. 
Using Sec.~\ref{sec:revOverhead}, we estimate that the related quantities for a reversible computation are $g_{\rm R} = 10^{19}$, $s_{\rm R} = g + s \approx 10^{19}$, and $t_{\rm R} = 10^{11}$.

We now turn to estimating a rough upper bound for the time and space overheads for {\sc Quall-E}, an instantiation of the partially reversible HLAI with each fully reversible segment run in a quantum coherent manner. 
We will estimate various quantities including $T_{\rm Q}$, the time to implement a segment on a quantum computer. We will be particularly interested in the ratio $\xfrac{T_{\rm Q}}{T}$, representing the speed of {\sc Quall-E}'s thought-processes relative to a human's.

\subsection{Logical quantum resources for {\sc Quall-E}}
\label{sec:logicalQResources}

The above estimates on the number of reversible operations $g_{\rm R}$, the logical circuit depth $t_{\rm R}$, and the number of bits $s_{\rm R}$ would, on an ideal quantum computer, translate directly to the number of logical quantum gates $g_{\rm I}$, the logical quantum circuit depth $t_{\rm I}$, and the number of logical qubits $s_{\rm I}$, with the releasing of ancilla bits becoming measurement and resetting of ancilla qubits. On an actual device, however, there will be compilation overhead, both from routing and gate synthesis. Physical constraints limit the connectivity achievable between logical qubits, \ie, on the locations of logical qubits in the quantum processor on which multi-qubit logical gates can be performed. Routing moves information around in the processor to enable the logical gates to be performed. The precise amount of connectivity differs from one quantum computing architecture to another. Large-scale quantum computers will integrate quantum and classical processors and have a modular structure with connections at various different scales \cite{bravyi2022future}.

Alon \etal~\cite{Alon94} show that any circuit on $s_{\rm I}$ qubits can be implemented on any connected architecture with gate routing overhead at most a multiplicative factor of $3s_{\rm I}$, taking the total logical gate count to no more than $3s_{\rm I}g_{\rm I}$. On some architectures one can do much better. Let’s consider an architecture in which each chip contains only a small number of logical qubits~\cite{Nic14}. or simplicity of analysis, we  will consider architectures made up of $s_{\rm I}$ chips, each with 1 logical data qubit and a couple of logical ancillae that can temporarily hold logical data qubits, routed from other chips, so that a Toffoli gate can be applied on chip to a triple of logical qubits. This overhead could clearly be reduced, but this factor of $3$ is small compared to other considerations in this analysis. Thus, we consider an architecture in which each of the $s_{\rm I}$ chips can hold $u = 3$ logical qubits, resulting in a total hardware requirement, in terms of logical qubits, of
\beq
s_{\rm L} = u s_{\rm I} = u(g + s).
\label{eqn:s_L}
\eeq
These chips would be connected via photonic networks that are not required to be planar, but with limitations on the connectivity, particularly the degree of the connectivity graph. 

There are a number of reasonable connectivity architectures. For a graph with the topology of an $D$-dimensional hypercube, in the case in which the number of nodes (chips in our case) is $n= 2^D$, Beals \etal~\cite{Bea13} show that any quantum circuit with $g$ gates on $n$ qubits can be implemented in depth increased by at most a multiplicative factor of $D(D+1)/2$. In our case, with $s_{\rm I} = 10^{19}$ chips, each containing one logical data qubit, $D = \log_2 s_{\rm I} \approx 64$, the depth would increase by a factor of $\sim 2000$. A better choice would be the cyclic butterfly graph, which shares many properties of a hypercube (it is a ``hypercubic’’ graph), but has constant degree and reduced routing overhead; Brierley~\cite{Bri15} shows that for the cyclic butterfly graph the depth increases by at most a factor of $6\log_2 n$, which in our case would be a factor of $\sim 384$. Herbert~\cite{Her18} conjectures that, for almost all $3$-regular graph architectures, the factor would be $4\log_2 n$, reducing the estimate to $\sim 256$.

We also need to include overhead from gate synthesis. Which logical gates are native to an architecture varies, and Toffoli gates are not usually native to the hardware. Thus there will be a small constant $c$ in overhead due to synthesis.
The exact value of the gate synthesis overhead depends on the gate set available. 
Altogether, using Herbert's conjecture, we estimate the overhead due to gate synthesis and routing to be 
$q = \rfrac{40}{3} c \log_{10}(s_{\rm I})$
(using $\log_{10}(2) \approx 3/10$).
For $s_{\rm I} = 10^{19}$, we have 
$q = \xfrac{760c}{3}$. For our estimates, we’ll use $q = 10^3$, so the logical depth increases from $t_{\rm I} = 10^{11}$ to $t_{\rm L}=10^{14}$. 
In terms of the input parameters, this is 
\beq 
t_{\rm L} = qt = 4ct \log_{2}(g).
\label{eqn:t_L}
%t_{\rm L} = qtT' = \rfrac{40}{3} ctT' \log_{10}(gT'). \label{eqn:t_L}
% t_{\rm L} = \rfrac{40}{3} c \log_{10}(gT') \frac{gT'}{k}. \label{El1}
\eeq
To summarise, we will use the rough upper bounds of  $t_{\rm L} = 10^{14}$ for the logical depth, and ${s}_{\rm L} = 3\times 10^{19}$ for the number of logical qubits, in our next set of estimates.

\subsection{Overhead due to error correction and fault tolerant gate times}
\label{sec:errCor}

We now turn to providing a rough upper bound on the actual time $T_{\rm Q}$ this computation would require, given that fault tolerant quantum gates are slow compared to classical gates. Babbush \etal~\cite{Bab21} estimate the time for a fault tolerant Toffoli gate 
as $\tau_{\rm LTof} = 170 \mu$s, using a distance $30$ surface code on a superconducting qubit device. This is about 7 orders of magnitude slower than doing a Toffoli gate on a standard classical processor. (Babbush \etal~give $330$ps, the typical speed of a $3$-GHz CPU, as their estimate for the time of a classical Toffoli gate, but mention that this is being generous to the quantum computer since generally thousands of Toffoli gates would be involved in a single operation on a classical computer, which would make it consistent with our estimate.) 
This is discouraging since, together with our depth estimate of $t_{\rm L} = 10^{14}$, we have 
$T_{\rm Q} = t_{\rm L}\tau_{\rm LTof} = 1.7\times 10^{10}$s. That is, {\sc Quall-E} would take over 500 years to have thoughts that would take only $1$s if he were a human. 

Our context, however, is quite different from the one Babbush \etal~consider, so we will carry out our own estimates for $\tau_{\rm LTof}$. Their estimate is for a small, fault tolerant device early in the fault tolerant era.  It is reasonable for us to posit lower error rates and higher parallelism than they did, which will reduce the resources required. On the other hand, the computations they used for their estimates involve $10^2$ to $10^9$ logical qubits and a similar number of logical gates, which are substantially smaller than ours. The required code distance is key to the resource estimates, and ultimately the time taken. There are tradeoffs here: because we require many more qubits and gates, we may need to encode in a higher-distance code, but on the other hand, a lower error rate would allow us to use a smaller-distance code.

To begin our estimate for a rough upper bound on the required code distance, which will help determine the physical resources in time and space required, we estimate the probability of an error-free run for a circuit of depth $2t_{\rm L}$, the depth of a segment of the HLAI run forward and in reverse. We consider the simple and common model of independent and identically distributed (i.i.d.) errors, in which there is a probability $p_{\rm L}$ of a logical Pauli error at each error location, \ie~at each logical qubit after each time step of the algorithm. The number of error locations $\ell = 2t_{\rm L}s_{\rm L}$ is the product of the logical depth and number of logical qubits.
Under this model, the probability of
at least one error in the run is $1 - (1 - p_{\rm L})^\ell$.
Our estimates of depth of $2 t_{\rm L} = 2\times 10^{14}$ and number of qubits $s_{\rm L}=10^{19}$ yield $\ell \approx 10^{33}$. 
We'll take as our acceptable probability of any logical errors occurring in a run to be $\epsilon=10^{-2}$.
Thus, to first order, we would need the logical error rate $p_{\rm L}$ to be no more than on the order of $p_{\rm L}=\epsilon/\ell = 10^{-35}$. 

Fowler \etal~\cite{FMM12} estimate that for the surface code $p_{\rm L} \approx 0.03 ({p}/{p_{\rm th}})^r$, where $p_{\rm th}$ is the fault tolerance threshold for the code, $p$ is the physical error rate for a simple error model at the physical level, and $r = \frac{d + 1}{2}$, where $d$ is the code distance. 
Taking $p_{\rm L} = (2 t_{\rm L}s_{\rm L})^{-1}$, we require distance $d = \lceil 2r - 1\rceil
$ where 
\beq 
r = \log_{\xfrac{p}{p_{\rm th}}}\left(\frac{100\epsilon}{6t_{\rm L}s_{\rm L}}\right).
\label{eqn:r}
% = \frac{\log_{10}(\xfrac{2t_{\rm L}s_{\rm L}}{0.3})}{\log_{10}(\xfrac{p_{\rm th}}{p})}.
\eeq 
If we took the physical error rate $p$ to be approximately $0.1 p_{\rm th}$, as Babbush \etal~do, we would need distance $d \approx 66$ to obtain $p_{\rm L} < 10^{-35}$, more than double their code distance of $30$. Our experiment would require us to be further into the fault tolerant era, so it seems reasonable to hope for a physical error rate an order of magnitude smaller, $p = 0.01 p_{\rm th}$.  For that, we would need a $d  \approx 34$ surface code. Thus, the effects from the improved error rate and the much larger number of logical qubits roughly cancel each other out. The physical resources corresponding to this estimate is in line with a common rule-of-thumb estimate of $1000$ physical qubits per logical qubit. 

Babbush \etal\ estimate $\tau_{\rm QEC} = 1 \mu$s, where $\tau_{\rm QEC}$ is the time taken for one round of surface code error correction including gate operations, measurement, and decoding. Their estimate of $170 \mu$s for a logical Toffoli comes from multiplying by $5.5 d$, the number of surface code cycles they estimate are needed per logical Toffoli gate, using a code distance of $d=30$. They give a detailed analysis of an approach using Toffoli factories to distill Toffoli gates, concluding with the $5.5 d$ estimate.
Altogether, we have as our estimate
\beq
\tau_{\rm LTof} = 5.5d \tau_{\rm QEC}.
\label{eqn:tau_LTof}
\eeq
Keeping their estimate for $\tau_{\rm QEC}$ and since we are using the slightly larger code distance of 34, our time would go up to $189 \mu {\rm s}$. Babbush \etal\ estimate a single Toffoli factory needs $144 d^2$ physical qubits, or
$\sim 1.6\times 10^5$ in our case.

To support our $10$ million concurrent processes as discussed in Sec.~\ref{sec:segmentedModel}, we would need $10$ million such factories. Altogether that gives a total of roughly $1.6\times 10^{12}$ physical qubits needed to implement the Toffoli factories, a tiny fraction of the total number of physical qubits used to implement the $10^{19}$ logical qubits. Babbush \etal\ are able to ignore routing time because their model has only a single gate happening at each time step, which can be routed with a small constant overhead~\cite{BVE21}. We need to include the output of the $10$ million Toffoli state factories in our routing, but since again that is a small fraction compared to the total number of logical qubits, our previous estimate suffices, leaving us with an estimate of $189\mu$s per logical Toffoli gate.  Together with our depth estimate of $t_{\rm L} = 10^{14}$, this yields an estimate of $T_{\rm Q}$ of $1.89\times 10^{10}{\rm s}$  (which is almost $600$ years). 

One way to reduce the fault tolerant overhead would be to take a different approach to implementing fault tolerant Toffoli gates. Purely 2D approaches do not support fault tolerant Toffoli gates, or more generally a universal set of fault tolerant gates, and magic state factories of one sort or another are a common approach to handling this deficiency. On the other hand, in 3D, there are sets of transversal gates that are universal. Bombin~\cite{Bom18} recognised that one can relax the 2D approach, so that it mimics a 3D code playing out over time, with decoding happening across time steps as sufficient information comes in. Brown~\cite{Bro20} builds on Bombin’s approach to exhibit a transversal Toffoli gate for the surface code that requires $3d$ surface code cycles, where $d$ is the distance of the code, reducing the logical Toffoli gate time to $99 \mu {\rm s}$ for distance-$33$ codes. 
Clearly we will need more radical advances to bring $T_{\rm Q}$ down to human conversational scales.

\subsection{Potential ways forward}
\label{sec:wayForward}

Many advances are expected in quantum computing in the coming decades, and much remains unknown about the structure and resource requirements for an HLAI. Here, we discuss the potential impact advances would have in bringing down our rough upper bound estimates to enable {\sc Quall-E} to interact at human speeds.
Altogether, from equations (\ref{eqn:s_L}) and (\ref{eqn:t_L}) in Sec.~\ref{sec:logicalQResources} and equations (\ref{eqn:r}) and (\ref{eqn:tau_LTof}) in ~\ref{sec:errCor}, we have, as our upper bound estimation for the {\sc Quall-E} to human ratio of times required for a given thought-process,  
\beq
\begin{split}
\frac{T_{\rm Q}}{T} &= 
t_{\rm L}\tau_{\rm LTof} \\
&=\frac{22caF}{kb} \log_{2}(aFT)
d
\tau_{\rm QEC},
\end{split}
\label{EqTimeRatio}
\eeq
where 
\beq
d = \left\lceil 2\log_{\xfrac{p}{p_{\rm th}}}
\left(\frac{25\epsilon}{6uctT(aFT + S)
\log_{2}(aFT)}\right)- 1\right\rceil.
\eeq
Here we are using input (as opposed to derived) parameters; as a reminder, $F$ and $S$ are the input estimates of FLOP/s and bits for an HLAI, 
$T$ is the input estimate for a human to make an observation, contemplate it, and compose a message, $a$ is the conversion factor between floating point operations and Boolean operations, $b$ is the Boolean parallelization factor, $c$ is the gate synthesis overhead factor,  $\epsilon$ is the acceptable probability of a segment having one or more errors, $k$ is the amount of parallelism, $\xfrac{p}{p_{\rm th}}$ is the ratio of the physical error to the threshold error. The equation assumes surface code encoding and the architecture described above.
Looking at this equation, we see that the quantities that have the most impact are $\tau_{\rm QEC}$ and the circuit depth per unit time, $\delta = \xfrac{aF}{kb}$, of a classical HLAI (from Section \ref{sec:hlai_estimates}).

We first turn to advances that could bring down $\tau_{\rm QEC}$, the time for one round of error correction. It depends on the physical two-qubit gate time, measurement time, gate synthesis time, and decoding time. Bringing down these times are a high priority for realising quantum computers generally, and are active areas of research. 

In terms of hardware improvements, many groups around the world are working on a variety of different hardware approaches to achieve lower error rates, faster measurements, and faster gate speeds. It seems reasonable to expect that physical gate speeds and measurement speeds will improve. 
Advances in control within the last couple of decades have been impressive, and there are many possible physical systems that could be used as qubits once they can be sufficiently controlled. As an example, a high-fidelity $0.8$ns two-qubit gate was demonstrated in phosphorous-donors in silicon~\cite{he2019two}, though the short-range nature of the interaction means it will be challenging to scale up to a large-scale device~\cite{chatterjee2021semiconductor}. We can take inspiration from classical computer clock speeds, which improved three orders of magnitude in less than three decades (from $1$ MHz in the early 1970s to $1$ GHz in the late 1990s~\cite{clockSpeeds}), and from a 2022 demonstration of a femtosecond ($10^{-15}$s) classical logic gate~\cite{femtoClassicalGate}. 
Back on the quantum side, breakthroughs in optical frequency comb control now enable atomic clocks to operate at optical frequencies~\cite{nist_optCombs, Cundiff_optCombs} (beyond $10^{15}$Hz).  
Further on the quantum side, 2022 also saw the demonstration of a Rabi cycle (that is, a single-qubit gate) taking only 52fs, using extreme-ultraviolet pulses acting on Helium atoms~\cite{Nandi2022}.

Our estimates already postulated an order of magnitude decrease in error rates. Further lowering the error rate would help, but because $T_{\rm Q}$ depends only logarithmically on the error rate, the improvements would have to be by multiple orders of magnitude to have a significant impact on the time (though smaller improvement would have a more significant impact on space requirements). For example, an additional order of magnitude ($p = 0.001 p_{\rm th}$) would reduce the required code distance to $d = 22$, having only a small effect on $\tau_{\rm LTof}$, 
but more than halving the number of physical qubits per logical qubits. 

In addition to hardware improvements, we can expect that improvements can be made in decoding, in error correction more generally, and in fault tolerant implementations. These are again active areas of research in which significant advances continue to be made at a rapid pace. Just within the time that this paper was being finalised, there were advances in decoding, including a pre-decoding for surface codes that brings down the overhead by multiple orders of magnitude~\cite{smith2022local} and a logarithmic parallel decoder for asymptotically good LDPC codes~\cite{leverrier2022parallel}, quantum Tanner codes which themselves are a recent innovation~\cite{panteleev2022asymptotically,leverrier2022quantum}. Some LDPC codes support single-shot error correction, which reduces the error correction cycles from $O(d)$ to $O(1)$, immediately improving the depth by more than an order of magnitude in our case. Powerful generalisations of error correction approaches have also been developed within the last year or so, including dynamical logical qubits~\cite{hastings2021dynamically}.  

It seems plausible that a combination of advances in error correction and fault tolerance, including decoding, and improved speeds for 2-qubit gates and measurements, could bring our estimates down by 4 to 6 orders of magnitude, possibly more. Advances need to happen on multiple fronts to achieve this. For example, without improvements in gate and measurement speeds, more efficient error correction would be limited to 2 to 3 orders of magnitude because the logical Toffoli gate speed is lower bounded by the physical gate speed. On the other hand, if the physical gates got a lot faster, the decoding time would become a bottleneck if it could not be improved. 

Another reason we would expect $T_Q/T$ to come down is that once a specific HLAI algorithm is discovered, it is likely that we can make use of its structure to reduce both the reversibility overhead and the routing overhead. While Bennett’s reversibility construction works for any possible algorithm, in practice there are usually better ways to obtain a reversible, and therefore a quantum, version of a classical algorithm (e.g.~\cite{bhaskar2015quantum,hadfield2018quantum,parent2015reversible,haner2018optimizing}). Furthermore, our estimates for routing likely substantially overestimate the time needed. The estimates support routing of qubits furthest away from each other at each step. Not only is it generally unlikely that the maximum routing distance is needed at each step, but given the suspected highly parallel nature of an HLAI algorithm, with separate subroutines running and only loosely interacting, it is likely that an architecture well matched to the structure could keep most of the routing within sets of qubits participating in a single subroutine, with only occasional need for larger scale routing. It seems likely that taking advantage of the structure could bring down the routing overhead by an order of magnitude and the reversibility overhead by another 2 to 3 orders of magnitude, and plausible that clever use of the structure could get us to 
beyond these 4 
orders of magnitude of improvement. 

There are other speculative possibilities that could help to approach human speeds. Once the structure of an HLAI algorithm is known, we may be able to take advantage of that structure in other ways. It seems likely that much of {\sc Quall-E}’s thinking during the experiment would not be influenced by the observation given that the number of floating-point operations is the same order of magnitude as the number of bits and the hypothesised extremely parallel nature of the HLAI algorithm. Or, somewhat along similar lines, {\sc Quall-E}’s intense concentration on the observation could mean that some of the AI processes would not be happening at all during the experiment. Either of these possibilities could significantly reduce the amount of computation that needs to be implemented in a way that can be reversed, thus reducing quantum computational resource requirements. 

Another possibility would be that the HLAI algorithm has inherent robustness and thus requires less overhead in terms of error correction and fault tolerance. Another approach to reducing the time and resources in a quantum implementation would be to use quantum algorithms to speed up subprocesses, with the expectation that square-root speed ups may be reasonable (\eg~\cite{AmbKok17, Mon20, BOJ21}). This use of quantum algorithms would be unusual in that it would not be using them for quantum advantage, but rather to make up for the large constant and logarithmic factors that mean that classical algorithms when implemented on a quantum computer run significantly slower than on classical computers. A related speculative possibility is the development of efficient ``native-quantum” HLAIs for which there isn’t a classical analogue, in which case there is of course no overhead due to a conversion from classical to quantum. Such possibilities are left for future investigation.

Given the large variation in estimates for the resources needed to implement a HLAI, it seems worthwhile to walk through a few possible scenarios. If the higher estimates of hardware and error correction advances are realised, improving $\tau_{\rm LTof}$ by 6 orders of magnitude and taking advantage of structure provides another 4 orders of magnitude, those alone remove the $10^{10}$ overhead from our original estimates, bringing {\sc Quall-E} to human speed. If instead only the lower projections of 4 orders of magnitude for error correction and 3 for taking advantage of structure were realised, human speeds would be attainable if, for example, either an HLAI could be implemented with 3 orders of magnitude fewer operations than the estimate we used, or if less of the HLAI computation would need to be reversed. 

Regarding the possibility that 3 orders of magnitude fewer operations are needed for HLAI than $10^{15}$ FLOP/s we used as input for our estimates, it is notable that while Carlsmith thought this would likely be sufficient, (as we quoted earlier), he also finds $10^{13}$ or even $10^{12}$ plausible~\cite{car20}. Moreover, ChatGPT3, which displays aspects of human-level intelligence, would require on order $10^{12}$ FLOP/s to generate output at human speed (as opposed to the faster-than-human speed at which it normally generates answers). We base this claim on the following~\cite{Christiano23} ``GPT-3 [has] about 2$\times10^{11}$ parameters and uses about 4 flops per parameter per token [generated], so about $\times10^{12}$ flops per token. If a human writes at 1 token per second, then you should be comparing [that to] $10^{12}$ flops \dots per second.'' Here a ``token'' is a word generated by a Large Language Model. For the figure of roughly 4 flops per parameter per token, see Table~D.1 of~\cite{brown2020language_arXiv}\footnote{There is a published version with the same title~\cite{brown2020language}, but, unconscionably, it is significantly shorter and does not contain the relevant table.}. 
Changing the initial estimate to $10^{12}$ FLOP/s would bring down the depth (and hence time) directly by a factor of $10^3$, but not much more. That is because the smaller computation size (roughly $10^{16}$ Toffoli gates) would only have a small effect on $\tau_{\rm LTof}$: using $\tau_{\rm QEC}=1\mu$s as previously   would yield $\tau_{\rm LTof} =126\mu {\rm s}$.  

On the other hand, $10^{15}$ FLOP/s could turn out to be too low. While Carlsmith estimates that it is unlikely ($<10\%$) that more than $10^{21}$ FLOP/s would be needed, this statement suggests to us that it would be unwise to rule out such a high figure. Note that a rate of $10^{21}$ FLOP/s is 3 orders of magnitude beyond the capabilities of today's fastest supercomputer. Even if that rate is required, it is still possible that a combination of technical advances, taking advantage of structure to decrease not only the routing and reversibility, but also the amount that needs to be reversed in the first place, and the possibility of using quantum subroutines, could enable a {\sc Quall-E} able to perform the experiment and interact at human speeds. However, this eventuality would likely push the experiment considerably further into the future.

\section{Conclusion} \label{sec:disc}

Building on the Local Friendliness no-go theorem of Bong \etal~\cite{CQD20}, we have presented a new theorem establishing a contradiction between four metaphysical assumptions (one now actually called {\sc Friendliness}) and two technological assumptions. None of the metaphysical assumptions mentions quantum theory. While ``quantum'' obviously appears in the technological assumption of {\sc Universal Quantum Computing}, it still does not assume the universal validity of QT. It is merely an assumption about what is technologically possible, guided by the best operational theory we have, QT. Thus the theorem is of precisely the same sort as Bell's theorem. It is a strict theorem of experimental metaphysics. If the technology can be realised and performs as expected (and this is, just as it once was in Bell's case, a scientific as well as an engineering question) then one of the four metaphysical assumptions must be wrong. 

For each of the six assumptions, there is, amongst the plethora of approaches (interpretations and modifications) of QT, at least one that rejects that assumption, and arguably only that one~\cite{WisCav22}. This shows the independence and relevance of all the assumptions. But what about QT itself? What if one were to believe the claim that \guillemotleft Quantum Theory Needs No ``Interpretation''\guillemotright~\cite{FucPer2000}, and to follow Fuchs (pre-QBism) and Peres in neither modifying nor interpreting QT? The answer, by our new Theorem, is that one would be mistaken and misled. 
One has to give up at least one of the six widely held assumptions. It is only by carefully interpreting, or modifying, QT, that it can be made compatible with the theorem while remaining consistent with existing experiments and experiences.  

The final undertaking of this paper was to estimate the level of technology required to perform an experiment of the type relevant for our theorem. We found that this level is far in advance of current technology, both in scale and rate of quantum information processing. It is thus of interest to identify meaningful experimental milestones that could be achieved between now and when an HLAI could be implemented on a UQC, along the lines of Ref.~\cite{CQD20} but with ever more sophisticated information processors in the role of the “friend”. This is a problem for future work,
one that could be helped by software such as Quanundrum \cite{nurgalieva2022thought}.

As one example of a nearer-term milestone, the single physical qubit of Ref.~\cite{CQD20} could be replaced by a logical qubit encoded in an error-detecting or error-correcting code, using a  small number of physical qubits. Alternatively, it may also be possible to implement some minimal degree of agent-like behaviour with a relatively small number of physical qubits. In the decades before the full experiment proposed here can be realised, there is also much theoretical work to be done, in formulating suitable metaphysical assumptions for such intermediate experiments to have new and non-trivial metaphysical implications. \break

\begin{acknowledgments}
The paper has benefited from astute feedback by Nora Tischler. EGR thanks Forrest Bennett and Steve Omohundro for pointers to estimates for AI memory and processing speed requirements. We thank Y\`il\`e Y{\=\i}ng for help in preparing the paper, and George Musser Jr for probing questions. HMW and EGC acknowledge the traditional owners of the land at Griffith University on which this work was undertaken, the Yuggera and Yugambeh peoples. 

Following the appearance of version 1 of this article on the arxiv, the authors are grateful for comments made,  at workshops on this topic which we co-organised, 
by Mohan Sarovar, Marissa Giustina, Lidia del Rio, and, most especially, Richard Healey, who provided a very detailed critique. We also thank Paul Christiano for information on the processing requirements of running ChatGPT. We thank Y\`il\`e Y{\=\i}ng for further help in preparation and for insightful comments. 

This research was supported by grant number FQXi-RFP-CPW-2019 from the Foundational Questions Institute and Fetzer Franklin Fund, a donor advised fund of Silicon Valley Community Foundation (HMW and EGC), by the Australian Research Council Centre of Excellence Program CE170100012, the Centre for Quantum Computation and Communication Technology (CQC2T) (HMW), and by the Australian Research Council (ARC) Future Fellowship FT180100317 (EGC).  

EGR is grateful for support from the NASA Ames Research Center, from the NASA SCaN program, and from DARPA under IAA 8839, Annex 130, and for the collaborative agreement between NASA and CQC2T.

Summary of contributions: The paper was conceived by HMW, who wrote a first draft. All three authors contributed substantially to the first four sections. HMW and EGC were the primary contributors to Section 5. EGR was the lead contributor to Section 6.
\end{acknowledgments}\break

\begin{appendix}
\section{Deutsch's contributions and $\ket{\text{``I knew 0 or 1''}}$}
\label{App:Deutsch}

The first general and fully quantum model of computation was famously introduced by Deutsch in 1985 in  Ref.~\cite{Deutsch85b}, whose very title introduced the term ``Universal Quantum Computing''. But it was a somewhat less famous paper by him~\cite{Deutsch85a}, published in the same year but submitted some months earlier, which had already  suggested the idea of applying such a quantum computer (albeit not by that name) to the Wigner's friend scenario. Deutsch begins by asserting that our assumption 5, {\sc Human-Level Artificial Intelligence} is true:
\begin{quote}
    Sooner or later (Turing, 1950) there will be machines
capable of independent thought comparable in every way to that of human
beings. One of them could no doubt be persuaded to take part in this 
experiment. 
\end{quote}
As this quote shows, he even touches upon the ethical issue of consent, which we considered in Sec.~\ref{subsec:MethodDesutch}. Next, he discusses the physical construction of a computer suitable for exactly the task we have discussed above:
\begin{quote}
    Presumably its internal workings will be electronic, rather than
biological, and its \dots\ Hamiltonian \dots\ will be known to
the designers. Extra apparatuses would have to be installed to give it the
requisite sense organ [to measure the qubit], and to allow the total Hamiltonian to be temporarily
altered when necessary. Sufficient coherence for the interference effects to be preserved will be possible if, for example, the information in the sense organ, the memory, and all other affected parts of the observer are stored in sufficiently microscopic finite-state components, thermally isolated from the
outside world. Another possibility might be to replace all the components
by logically equivalent systems of currents in superconductors.
\end{quote}
It is amusing to note that superconducting qubits are the current state of the art for quantum computers, but few could have predicted that in 1985. 
 
Now in this paper Deutsch also considers adding another subsystem, with which the HLAI (whom we are calling {\sc Quall-E} and whom Deutsch calls simply ``the observer'' or ``Professor X'') can interact, so that\footnote{Here and below we have made minor changes in the notation and terminology (remember that this was ten years before the word ``qubit'' was introduced~\cite{Schumacher}) to accord to that of the current paper.}  
\begin{quote}
After the completion of the measurement, the
observer records (in his memory, or in his notebook if necessary) --- not the
value ``0'' or ``1'' of the bit, but only whether or not he knows this value.
He may write ``I, Professor X, F.R.S., hereby certify that at time $t'''$, I have
determined whether the value of the qubit in the logical basis is 0 or 1. At this moment I am contemplating in my own mind one,
and only one of those two values. In order to facilitate the second part of
this experiment, I shall not reveal which one.'' This constitutes a record of
the completion of the measurement, a record which, we shall see, need not
be destroyed by a subsequent interference experiment.
\end{quote}
 That is, Deutsch is saying that the measurement reversal, \erf{undomeas}, can leave the state of this subsystem (be it some part of {\sc Quall-E}'s internal memory or an external notepad) unchanged. In our notation, the final state would be 
 \beq \label{keepflag}
 \sum_{b=0,1} c_b \ket{b} \otimes \ket{{\text{\sc Quall-E}}_{\rm ready}}\otimes \ket{\textrm{``knew 0 or 1''}},
\eeq
still leaving the qubit in its initial state to allow Alice to measure it. 
 The point of this is, for Deutsch, that (his emphasis) 
 \begin{quote}
    \emph{[T]he record that the 0/1 value of the bit was known to the
observer at time $t'''$ is preserved.} \dots\  The interference phenomenon
seen by our observer at the end of the experiment requires the presence of
both bit values, though he accurately remembers having known at a
previous time that only one of them was present. He must infer that there
was more than one copy of himself (and the qubit) in existence at that time,
and that these copies merged to form his present self.
 \end{quote}
That is, the experiment would establish, as far as Deutsch is concerned, Quantum Theory as a Universal Physical Theory (the title of his paper), and that the Everett interpretation is the only viable one. 

We discuss Everett's interpretation and several other approaches to quantum theory in Sec.~\ref{subec:RelInt}. Here, we wish to comment on the idea of the preservation of {\sc Quall-E}'s memory, or his record, of having seen {\em some} definite result. This idea was very influential on Brukner, whose work~\cite{BruknerLF} inspired the LF no-go theorem. In that work, Brukner sees this as the key point of Ref.~\cite{Deutsch85a} (with the friend here being feminine): ``The novelty of Deutsch’s proposal lies in the possibility for Wigner to acquire direct knowledge on whether the friend has observed a definite outcome upon her measurement or not
without revealing what outcome she has observed.'' As we have seen, Deutsch actually emphasized the ability of the ``observer''  (\ie~the ``friend'' in Brukner's parlance) to ``remember'' \emph{himself} that he had made an observation, but the distinction is not so important. 

Deutsch's proof that this is mathematically possible is certainly correct. Indeed, it is trivially so, since all that is required is an independent unitary evolution for the final subsystem over the course of the experiment: 
\beq \label{sillyflag}
 \ket{\textrm{blank}} \to \ket{\textrm{``knew 0 or 1''}},
 \eeq
Of course such a manner of producing this output would be silly, because it would undermine the motivation for this being a convincing \emph{record} of {\sc Quall-E} having known, and thought about, the bit value. It would be a meaningless, mechanically produced, sequence of ink stains on a notepad, unrelated to anything {\sc Quall-E} actually thought. But, we claim, this is the  best {\sc Quall-E} can do: to ``press a button'' which automatically produces one particular message regardless of what he actually experiences.  

To see why this is the case, recall Deutsch's colourful description above. If Professor X F.R.S. truly was ``contemplating in [his] own mind one, and only one'' of the values 0 and 1, would those different contemplative experiences be consistent with his two different states (\eg~on the right-hand-side of  \erf{undomeas}) making {\em exactly} the same ink stains on his notepad, down to the atomic level and beyond? We think obviously no; they would preclude it. Even if the record were a far more constrained system, such as a finite string of bits, different ``contemplations'' would surely affect any genuine record of thoughts. For example, if {\sc Quall-E} were contemplating 0, he might, as he extemporizes this record for posterity, omit his postnomial, or perhaps put a comma after ``and only one''. Either would be more than enough to make the entire experiment fail, because the reversal operation, acting only on the qubit and {\sc Quall-E} subsystems, would not return these to their initial factorized state. The reversal would only work if {\sc Quall-E} committed, before the experiment starts, to creating a particular message at a particular time as in \erf{sillyflag}, which is thus no witness to {\sc Quall-E}'s real thoughts. 

Now, it may be possible to construct a device (another enormous quantum computer) that could perform a measurement on {\sc Quall-E}'s logical qubits that would distinguish that he was contemplating 0 or 1 (without distinguishing between these two), and raise a flag if he was. That would arguably be more convincing than an automatically produced flag. However, in order to trust such a flag, one would need to assume that the quantum description of the experiment is correct. But if one makes that assumption then the flag does not really serve much purpose. It is for these reasons that we did not (in Ref.~\cite{CQD20}), and do not here, consider such a record in our Wigner's friend scenario. 
\end{appendix}
\break

\bibliography{
QMCrefsPLUS.bib
}

\end{document}